\documentclass[10pt,twocolumn,twoside]{IEEEtran}
\renewcommand{\baselinestretch}{0.95}

\usepackage{color}
\usepackage{theorem}
\usepackage{times,amsmath,epsfig}
\usepackage{amssymb}
\usepackage{subfigure}
\usepackage{cite}
\usepackage{hyperref}

\usepackage{algorithmic}
\usepackage{algorithm}
\usepackage{needspace}

\newcommand{\myparagraphtc}[1]{\needspace{1\baselineskip}\medskip\noindent {\it #1.}\addcontentsline{toc}{subsubsection}{\qquad\qquad\quad#1}}

\usepackage{tikz}
\usetikzlibrary{shapes,arrows}
\input{mysymbol.sty}
\tikzstyle{phantom vertex} = [ ellipse, 
                               anchor = center, 
                               minimum height = 1*\unit, 
                               minimum width  = 1*\unit,
                               inner sep=0pt,
                               anchor=center]
\tikzstyle{red vertex}   = [black, fill = red!20,   phantom vertex, draw]
\tikzstyle{black vertex} = [black, fill = black!20, phantom vertex, draw]
\tikzstyle{blue vertex}  = [black, fill = blue!20,  phantom vertex, draw]
\tikzstyle{green vertex} = [black, fill = green!20,  phantom vertex, draw]
\tikzstyle{yellow vertex} = [black, fill = yellow!20,  phantom vertex, draw]
\tikzstyle{cyan vertex} = [black, fill = cyan!20,  phantom vertex, draw]
\tikzstyle{vertex}       = [draw, phantom vertex]

\tikzstyle{point} = [ellipse, inner sep=0pt, draw, fill=white, anchor = center,
                     minimum height = 0.05*\unit, minimum width  = 0.05*\unit]

\newtheorem{mytheorem}{\bf Theorem}

\newtheorem{myproposition}{\bf Proposition}
\newtheorem{remark}{\bf Remark}

\newtheorem{problem}{\bf Problem}

\title{\!Network\! Topology\! Inference\! from\! Spectral\! Templates}

\author{\IEEEauthorblockN{Santiago Segarra, Antonio G. Marques, Gonzalo Mateos, and Alejandro Ribeiro}
\thanks{Work in this paper is supported by Spanish MINECO grant No TEC2013-
41604-R and USA NSF CCF-1217963. S. Segarra and A. Ribeiro are with the Dept. of Electrical and Systems Eng., Univ. of Pennsylvania.  A. G. Marques is with the Dept. of Signal Theory and Comms., King Juan Carlos Univ. G. Mateos is with the Dept. of Electrical and Computer Eng., Univ. of Rochester.  Emails: ssegarra@seas.upenn.edu, antonio.garcia.marques@urjc.es, gmateosb@ece.rochester.edu and aribeiro@seas.upenn.edu. Part of the results in this paper were submitted to the \textit{2016 IEEE SSP Workshop}~\cite{SSAMGMAR_ssp16} and the \textit{2016 Asilomar Conference on Signals, Systems and Computers}~\cite{SSAMGMAR_Asilomar16}.}}

\begin{document}
\maketitle

\begin{abstract}%
We address the problem of identifying a graph structure from the observation of signals defined on its nodes.
Fundamentally, the unknown graph encodes direct relationships between signal elements, which we aim to recover from observable indirect relationships generated by a diffusion process on the graph. 
The fresh look advocated here permeates benefits from convex optimization and stationarity of graph signals, in order to identify the graph shift operator (a matrix representation of the graph) given only its \textit{eigenvectors}. These \emph{spectral templates} can be obtained, e.g., from the sample covariance of independent graph signals diffused on the sought network. 
The novel idea is to find a graph shift that, while being consistent with the provided spectral information, endows the network with certain desired properties such as sparsity. To that end we develop efficient inference algorithms stemming from provably-tight convex relaxations of natural nonconvex criteria, particularizing the results for two shifts: the adjacency matrix and the normalized Laplacian. 
Algorithms and theoretical recovery conditions are developed not only when the templates are perfectly known, but also when the eigenvectors are noisy or when only a subset of them are given.  
Numerical tests showcase the effectiveness of the proposed algorithms in recovering social, brain, and amino-acid networks.
\end{abstract}




\section{Introduction}\label{S:Introduction}


Advancing a holistic theory of networks necessitates fundamental breakthroughs in modeling,
identification, and controllability of distributed network processes -- often conceptualized as \emph{signals defined on the vertices of a graph}~\cite{barrat2012book,kolaczyk2009book}.
Under the assumption that the signal properties are related to the topology of the graph where they are supported, the goal of graph signal processing (GSP) is to develop algorithms that fruitfully leverage this relational structure~\cite{EmergingFieldGSP,SandryMouraSPG_TSP13}.
Instrumental to that end is the so-termed graph-shift operator (GSO)~\cite{SandryMouraSPG_TSP13}, a matrix capturing the graph's local topology and whose eigenbasis is central to defining graph Fourier transforms~\cite{SandryMouraSPG_TSP14Freq}. Most GSP works assume that the GSO (hence the graph) is known, and then analyze how the algebraic and spectral characteristics of the GSO impact the properties of the signals and filters defined on such a graph. Here instead we take the reverse path and investigate how to use information available from graph signals to infer the underlying graph topology; see also~\cite{DongLaplacianLearning,MeiGraphStructure,SSAMGMAR_ssp16,pasdeloup2016inferenceTSIPN16,Kalofolias2016inference_smoothAISTATS16}. 

Our focus in this paper is on identifying graphs that explain the structure of a random signal, meaning that there exists a diffusion process in the GSO that can generate the observed signal. Alternatively, we can say that the goal is to recover the GSO which encodes direct relationships between the elements of the signal from observable indirect relationships generated by a diffusion process. Such a problem is shown to be underdetermined and related to the concept of stationarity of graph signals~\cite{marques2016stationaryTSP16, perraudinstationary2016}. More precisely, it is established that the sought GSO must have the same eigenvectors as the signal's covariance matrix. This motivates a two-step network topology inference approach whereby we: i) leverage results from GSP theory to identify the GSO's eigenbasis from realizations of the diffused signal; and ii) rely on these (possibly imperfect and incomplete) \emph{spectral templates} to recover the GSO by estimating its eigenvalues. 

Network topology inference from a set of (graph-signal) observations is a prominent problem in Network Science~\cite{kolaczyk2009book,sporns2012book}. Since networks encode similarities between nodes, several approaches infer the so-termed \emph{association networks} by constructing graphs whose edge weights correspond to correlations or coherence measures indicating a nontrivial level of association between signal profiles at incident nodes~\cite[Ch. 7.3.1]{kolaczyk2009book}. 
This approach is not without merit and widely used in practice, but it exhibits several drawbacks, the main one being that links are formed taking into account only pairwise interactions, ignoring that the observed correlations can be due to latent network effects. Acknowledging these limitations, alternative methods rely on partial correlations~\cite{GLasso2008,kolaczyk2009book}, Gaussian graphical models~\cite{Lake10discoveringstructure, slawski2015estimation,meinshausen06,pavez_laplacian_inference_icassp16}, structural equation models~\cite{BazerqueGeneNetworks,BainganaInfoNetworks}, Granger causality~\cite{Brovelli04Granger,sporns2012book}, or their nonlinear (kernelized) variants~\cite{Karanikolas_icassp16,shen2016kernelsTSP16}.  Differently, recent GSP-based network inference frameworks postulate that the network exists as a latent underlying structure, and that observations are generated as a result of a network process defined in such graph. For instance, network structure is estimated in~\cite{MeiGraphStructure}  to unveil unknown relations among nodal time series adhering to an autoregressive model involving graph-filter dynamics. A factor analysis-based approach is put forth in~\cite{DongLaplacianLearning} to infer graph Laplacians, seeking that input graph signals are smooth over the learned topologies; see also~\cite{Kalofolias2016inference_smoothAISTATS16}. Different from~\cite{DongLaplacianLearning,MeiGraphStructure, Kalofolias2016inference_smoothAISTATS16} that operate on the graph domain, the goal here is to identify graphs that endow the given observations with desired spectral (frequency-domain) characteristics. Two works have recently explored this approach and addressed the problem of identifying a GSO based on its eigenvectors. One is \cite{SSAMGMAR_ssp16}, which assumes perfect knowledge of the spectral templates. The other is \cite{pasdeloup2016inferenceTSIPN16}, which only focuses on a Laplacian~GSO. 

After surveying the required GSP background, in Section \ref{S:prelim_problem} we formulate the problem of identifying a GSO that explains the fundamental structure of a random signal diffused on a graph. The novel idea is to search among all feasible networks for the one that endows the resulting graph-signal transforms with prescribed spectral properties (those guaranteeing graph stationarity~\cite{marques2016stationaryTSP16}), while the inferred graph also exhibits desirable structural characteristics such as sparsity or minimum-energy edge weights. It is argued that the required spectral templates can be pragmatically obtained, e.g., via principal component analysis (PCA) of an ensemble of graph signals resulting from network diffusion dynamics. Additional sources for the spectral templates are provided in Section \ref{Ss:remark_templates}. Using the templates as input, a fairly general optimization problem is then formulated to identify the network structure. For concreteness, emphasis is laid on the recovery of two particular GSOs; namely the adjacency matrix and the normalized graph Laplacian, but our methodology can be applied to other matrix representations of graphs. In Section \ref{Ss:size_feasible_set} we derive conditions under which the feasible set of the optimization problem reduces to a singleton, a situation in which pursuit of additional network structure is rendered vacuous. When multiple solutions exist, provably-tight convex relaxations -- leading to computationally-efficient algorithms -- are developed to identify the sparsest GSO consistent with the given eigenspace  (Section \ref{Ss:relaxation}). We then introduce an inference method for the pragmatic case where knowledge of the spectral templates is imperfect (Section \ref{Ss:ShiftRecImperfEig}), and establish that the proposed algorithm can identify the underlying network topology robustly. Last but not least, in Section \ref{Ss:ShiftRecIncompEig} we investigate the case where only a subset of the GSO's eigenvectors are known. Such incomplete spectral templates arise, for example, when the observed graph signals are bandlimited. Comprehensive numerical tests corroborate our theoretical findings and confirm that the novel approach compares favorably with respect to: (i)
established methods based on (partial) correlations; and (ii) recent graph signal processing-based topology inference algorithms (Section \ref{S:Simulations}). Test cases include the recovery of social and structural brain networks, as well as the identification of the structural properties of proteins from a mutual information graph of the co-variation between the constitutional amino-acids \cite{Marks2011proteins}.   

\noindent\emph{Notation:} The entries of a matrix $\mathbf{X}$ and a (column) vector $\mathbf{x}$ are denoted by $X_{ij}$ and $x_i$, respectively. Sets are represented by calligraphic capital letters and $\bbX_{\ccalI}$ denotes a submatrix of $\bbX$ formed by selecting the rows of $\bbX$ indexed by $\ccalI$. The notation $^T$ and $^\dag$ stands for transpose and pseudo-inverse, respectively; $\mathbf{0}$ and $\mathbf{1}$ refer to the all-zero and all-one vectors. For a vector $\bbx$, $\diag(\mathbf{x})$ is a diagonal matrix whose $i$th diagonal entry is $x_i$; when applied to a matrix, $\diag(\bbX)$ is a vector with the diagonal elements of $\bbX$. The operators $\circ$, $\otimes$, and $\odot$ stand for the Hadamard (elementwise), Kronecker, and Khatri-Rao (columnwise Kronecker) matrix products. $\| \bbX \|_p$ denotes the $\ell_p$ norm of the \textit{vectorized} form of $\bbX$, whereas $\| \bbX \|_{M(p)}$ is the matrix norm induced by the vector $\ell_p$ norm.


\section{Problem Statement}\label{S:prelim_problem}


A weighted and undirected graph $\ccalG$ consists of a node set $\ccalN$ of known cardinality $N$, an edge set $\ccalE$ of unordered pairs of elements in $\ccalN$, and edge weights $A_{ij}\in\reals$ such that $A_{ij}=A_{ji}\neq 0$ for all $(i,j)\in\ccalE$. The edge weights $A_{ij}$ are collected as entries of the symmetric adjacency matrix $\bbA$ and the node degrees in the diagonal matrix $\bbD:=\diag(\bbA\bbone)$. These are used to form the combinatorial Laplacian matrix $\bbL_c:=\bbD-\bbA$ and the normalized Laplacian $\bbL:=\bbI - \bbD^{-1/2} \bbA\bbD^{-1/2}$. More broadly, one can define a generic GSO $\bbS\in\reals^{N\times N}$ as any matrix having the same sparsity pattern of $\ccalG$~\cite{SandryMouraSPG_TSP13}. Although the choice of $\bbS$ can be adapted to the problem at hand, most existing works set it to either $\bbA$, $\bbL_c$, or $\bbL$. 

The main focus in this paper is on identifying graphs that explain the structure of a random signal. Formally, let $\bbx=[x_1,...,x_N]^T \in\mbR^N$ be a graph signal in which the $i$th element $x_i$ denotes the signal value at node $i$ of an unknown graph $\ccalG$ with shift operator $\bbS$. Further suppose that we are given a zero-mean white signal $\bbw$ with covariance matrix $\E{\bbw\bbw^T} = \bbI$. We say that the graph $\bbS$ represents the structure of the signal $\bbx$ if there exists a diffusion process in the GSO $\bbS$ that produces the signal $\bbx$ from the white signal $\bbw$, that is
\begin{align}\label{eqn_diffusion}
	\bbx\  =\ \alpha_0 \prod_{l=1}^{\infty} (\bbI-\alpha_l \bbS) \bbw
	\  =\ \sum _{l=0}^{\infty} \beta_l \bbS^l \bbw .
\end{align}
%
While $\bbS$ encodes only one-hop interactions, each successive application of the shift percolates (correlates) the original information across an iteratively increasing neighborhood; see e.g.~\cite{segarra2015graphfilteringTSP15}. 
The product and sum representations in \eqref{eqn_diffusion} are common -- and equivalent -- models for the generation of random signals. Indeed, any process that can be understood as the linear propagation of a white input through a static graph can be written in the form in \eqref{eqn_diffusion}. 
These include processes generated by graph filters with time-varying coefficients or those generated by the so-called \textit{diffusion} Laplacian \textit{kernels} \cite{smola2003kernels}, to name a few. 

The justification to say that $\bbS$ is the structure of $\bbx$ is that we can think of the edges of $\bbS$ as direct (one-hop) relationships between the elements of the signal. The diffusion described by \eqref{eqn_diffusion} generates indirect relationships. Our goal is to recover the fundamental relationships described by $\bbS$ from a set $\ccalX:=\{\bbx_p\}_{p=1}^P$ of $P$ independent samples of the random signal $\bbx$. 

We show next that this is an underdetermined problem closely related to the notion of stationary signals on graphs~\cite{marques2016stationaryTSP16}. Begin by assuming that the shift operator $\bbS$ is symmetric. Define then the eigenvector matrix $\bbV:=[\bbv_1,\ldots,\bbv_N]$ and the eigenvalue matrix $\bbLam:=\diag(\lam_1,\ldots,\lam_N)$ to write
\begin{align}\label{eqn_shift_operator_normal}
	\bbS =\bbV\bbLam\bbV^T.
\end{align}
Further observe that while the diffusion expressions in \eqref{eqn_diffusion} are polynomials on the GSO of possibly infinite degree, the Cayley-Hamilton theorem implies they are equivalent to polynomials of degree smaller than $N$. Upon defining the vector of coefficients $\bbh:=[h_0,\ldots,h_{L-1}]^T$ and the graph filter $\bbH\in\reals^{N\times N}$ as $\bbH:=\sum_{l=0}^{L-1} h_l \bbS^l$, the generative model in \eqref{eqn_diffusion} can be rewritten as
\begin{align}\label{E:Filter_input_output_time}
	\bbx  = \bigg(\sum_{l=0}^{L-1}h_l \bbS^l\bigg)\,\bbw
	= \bbH \bbw
\end{align}
for some particular $\bbh$ and $L$. Since a graph filter $\bbH$ is a polynomial on $\bbS$~\cite{SandryMouraSPG_TSP13}, graph filters are linear graph-signal operators that have the \textit{same eigenvectors} as the shift (i.e., the operators $\bbH$ and $\bbS$ commute). More important for the present paper, the filter representation in \eqref{E:Filter_input_output_time} can be used to show that \emph{the eigenvectors of $\bbS$ are also eigenvectors of the covariance matrix $\bbC_x:=\E{\bbx\bbx^T}$.} To that end, substitute \eqref{E:Filter_input_output_time} into the covariance matrix definition and use the fact that $\E{\bbw\bbw^T} = \bbI$ to write
\begin{align}\label{E:cov_output_filter}
	\bbC_x = \E{\bbH\bbw\big(\bbH\bbw\big)^T}
	= \bbH \E{\bbw\bbw^T}\bbH^T
	= \bbH\bbH^T .
\end{align}
If we further use the spectral decomposition of the shift in \eqref{eqn_shift_operator_normal} to express the filter as $\bbH = \sum_{l=0}^{L-1} h_l (\bbV\bbLam\bbV^T)^l =\bbV(\sum_{l=0}^{L-1} h_l \bbLam^l)\bbV^T$, we can write the covariance matrix as
\begin{align}\label{eqn_diagonalize_covariance}
	\bbC_x\  =\ \bbV\,\bigg|\sum_{l=0}^{L-1}h_l\bbLam^l\bigg|^2\,\bbV^T
	\ :=\ \bbV\diag(\bbp)\bbV^T,
\end{align}
where the matrix squared-modulus operator $|\cdot|^2$ should be understood entrywise, and we have defined the vector $\bbp:=\diag(|\sum_{l=0}^{L-1}h_l\bbLam^l|^2)$ in the second equality.

The expression in  \eqref{eqn_diagonalize_covariance} is precisely the requirement for a graph signal to be stationary \cite[Def. 3]{marques2016stationaryTSP16}; hence, the problem of identifying a GSO that explains the fundamental structure of $\bbx$ is equivalent to identifying a shift on which the signal $\bbx$ is stationary.
In this context, $\bbp$ is termed the power spectral density of the signal $\bbx$ with respect to $\bbS$. A consequence of this fact, which also follows directly from \eqref{eqn_diagonalize_covariance}, is that the \textit{eigenvectors} of the shift $\bbS$ and the covariance $\bbC_x$ are the same. Alternatively, one can say that the difference between $\bbC_x$, which includes indirect relationships between components, and $\bbS$, which includes exclusively direct relationships, is only on their \textit{eigenvalues}. While the diffusion in \eqref{eqn_diffusion} obscures the eigenvalues of $\bbS$, the eigenvectors $\bbV$ remain present in $\bbC_x$ as templates of the original spectrum.

Identity \eqref{eqn_diagonalize_covariance} also shows that the problem of finding a GSO that generates $\bbx$ from a white input $\bbw$ with unknown coefficients [cf. \eqref{eqn_diffusion}] is \emph{underdetermined}. As long as the matrices $\bbS$ and $\bbC_x$ have the same eigenvectors, filter coefficients that generate $\bbx$ through a diffusion process on $\bbS$ exist.\footnote{To simplify exposition, the general description of the recovery problem in this section assumes that neither $\bbS$ nor $\bbC_x$ have repeated eigenvalues. Technical modifications in the formulation to accommodate setups where the eigenvalues are not all distinct are discussed in Section \ref{Ss:ShiftRecIncompEig}.} In fact, the covariance matrix $\bbC_x$ itself is a GSO that can generate $\bbx$ through a diffusion process and so is the precision matrix $\bbC_x^{-1}$. To sort out this ambiguity, which amounts to selecting the eigenvalues of the shift, we assume that the GSO of interest is optimal in some sense. To be more precise, let $\ccalS$ be a convex set that specifies the type of shift operator we want to identify (details on $\ccalS$ are provided in Section \ref{Ss:AprioriInfoShift}) and let $\|\bbS\|_0$ count the number of nonzero entries in the GSO. We then want to identify $\bbS_0^*\in\ccalS$ with the smallest number of nonzero entries
\begin{alignat}{2}\label{eqn_zero_norm} 
	\bbS_0^* := &\argmin_{\{\bbS, \bblambda\}} \
	&&\|\bbS\|_0 ,                                    \nonumber\\ 
	&\text{s. to } && \bbS = \bbV\bbLam\bbV^T 
	= \sum_{k=1}^N \lambda_k\bbv_k \bbv_k^T,  \qquad
	\bbS \in \ccalS,
\end{alignat}
where $\bblambda=[\lambda_1,\ldots,\lambda_N]^T$. To simplify notation we have purposely ignored the optimal eigenvalues $\bblambda^*_0$ that belong to the argument of the minimum. Also, we have written $\bbV\bbLam\bbV^T \! =\! \sum_{k=1}^N \lambda_k\bbv_k \bbv_k^T$ to emphasize that if the eigenvectors $\bbv_k$ are known, the constraints in \eqref{eqn_zero_norm} are linear on the unknown eigenvalues $\lam_k$. Alternatively, we can introduce criteria in the form of generic convex functions $f ( \bbS, \bblambda)$ and define the shift operator that is optimal with respect to these criteria
\begin{alignat}{2}\label{E:general_problem} 
	\bbS^* := &\argmin_{\{\bbS, \bblambda\}} \
	&& f ( \bbS, \bblambda),                           \nonumber\\ 
	&\text{s. to } && \bbS = \bbV\bbLam\bbV^T 
	= \sum_{k=1}^N \lambda_k\bbv_k \bbv_k^T,  \qquad
	\bbS \in \ccalS.
\end{alignat}
Possible convex choices for the criteria in \eqref{E:general_problem} are to: (i) Adopt $f(\bbS, \bblambda) = f(\bbS) = \| \bbS \|_\mathrm{F}$ which finds a GSO that minimizes the total energy stored in the weights of the edges. (ii) Make $f ( \bbS, \bblambda) = f(\bbS) = \| \bbS \|_\infty$ which yields shifts $\bbS$ associated with graphs of uniformly low edge weights. This can be meaningful, e.g., when identifying graphs subject to capacity constraints. (iii) Minimize $f ( \bbS, \bblambda) = f(\bblambda) = -\lambda_2$, where $\lambda_2$ is the second smallest eigenvalue of $\bbS$. If the GSO is further assumed to be a Laplacian matrix, this yields a shift operator that promotes solutions with fast mixing times \cite{chung1997spectral}.

Independently of the criteria, the definitions in \eqref{eqn_zero_norm} and \eqref{E:general_problem} provide a formal description of a GSO $\bbS$ that is considered to be the best possible description of the structure of the signal $\bbx$. Our goal is to find estimators of these operators as described in the following two formal problem statements.

%
\begin{problem}\label{problem_known_matrix}
	Given a covariance matrix $\bbC_x$ identify the optimal description of the structure of $\bbx$ in the form of the graph-shift operator $\bbS_0^*$ defined in \eqref{eqn_zero_norm} or $\bbS^*$ defined in \eqref{E:general_problem}.
\end{problem}

%
\begin{problem}\label{problem_data_driven}
	Given a set $\ccalX:=\{\bbx_p\}_{p=1}^P$ of $P$ independent samples of the random signal $\bbx$ estimate the optimal description of the structure of $\bbx$ in the form of the graph-shift operator $\bbS_0^*$ defined in \eqref{eqn_zero_norm} or $\bbS^*$ defined in \eqref{E:general_problem}.
\end{problem}

%
Problem \ref{problem_known_matrix} is a simple convex optimization problem in the case of the convex objectives in \eqref{E:general_problem} but necessitates relaxations in the case of the minimum zero-norm formulations in \eqref{eqn_zero_norm}. To solve Problem \ref{problem_known_matrix} we use ensemble covariance matrices to obtain the eigenvectors and show that the estimation of the eigenvalues yields consistent estimators of sparse network structures. Problem \ref{problem_known_matrix} is addressed in Section \ref{S:ShiftInfFullEigen}. To solve Problem \ref{problem_data_driven} we first use independent samples of the random signal to estimate the covariance eigenvectors. Then we estimate the eigenvalues using reformulations of \eqref{eqn_zero_norm} and \eqref{E:general_problem} which are robust to errors stemming from the aforementioned eigenvector estimation step; see Section \ref{S:ShiftInfSubsetEigen} for a detailed treatment of Problem \ref{problem_data_driven}. Although Problem \ref{problem_known_matrix} can be thought as a prerequisite to study Problem \ref{problem_data_driven}, Section \ref{Ss:remark_templates} illustrates other situations in which the estimation of a GSO with prescribed eigenvectors is of practical interest. 

%
\begin{remark}[Precision matrices]\label{rmk_precision} \normalfont
	As already mentioned, the precision matrix $\bbC_x^{-1} \! = \! \bbV\bbLambda^{-1}\bbV^T$ is a possible solution to the problem of finding a GSO that explains the structure of $\bbx$. This establishes a clear connection between \eqref{eqn_zero_norm} and the problem of finding sparse estimates of precision matrices~\cite[Ch. 7]{kolaczyk2009book}. If the precision matrix $\bbC_x^{-1}$ is the {\it sparsest} matrix that explains the structure of $\bbx$, this matrix is also the solution to \eqref{eqn_zero_norm} and we have $\bbS_0^* = \bbC_x^{-1}$. In general, however, $\bbC_x^{-1}$ may not be sparse and, even if it is, there may be {\it sparser} graphs that explain  $\bbx$. In these cases the solution to \eqref{eqn_zero_norm} is a more parsimonious GSO. We can then think of \eqref{eqn_zero_norm} as a {\it generalization} of the problem of finding a sparse precision matrix.
\end{remark}

%
%

\subsection{A priori knowledge about the GSO}\label{Ss:AprioriInfoShift}

The constraint $\bbS \in \ccalS$ in \eqref{eqn_zero_norm}  and \eqref{E:general_problem} incorporates a priori knowledge about $\bbS$. 
If we let $\bbS = \bbA$ represent the adjacency matrix of an undirected graph with non-negative weights and no self-loops, we can explicitly write $\ccalS$ as follows
\begin{align} \label{E:SparseAdj_def_S}
	\ccalS_{\mathrm{A}} \!:= \! \{ \bbS \, | \, S_{ij} \geq 0, \;\,  \bbS\!\in\!\ccalM^N\!\!,\;\,  S_{ii} = 0, \;\, \textstyle\sum_j S_{j1} \! = \! 1 \}.
\end{align}
The first condition in $\ccalS_{\mathrm{A}}$ encodes the non-negativity of the weights whereas the second condition incorporates the fact that the unknown graph is undirected, hence, $\bbS$ must belong to the set $\ccalM^N$ of real and symmetric $N \! \times \! N$ matrices. The third condition encodes the absence of self-loops, thus, each diagonal entry of $\bbS$ must be null. Finally, the last condition fixes the scale of the admissible graphs by setting the weighted degree of the first node to $1$, and also rules out the trivial solution $\bbS\!=\!\bbzero$.  Naturally, the choice of the first node is (almost) arbitrary; any node with at least one neighbor in the sought graph suffices. Although not considered here, additional sources of information such as knowing the existence (or not) of particular edges can be incorporated into $\ccalS$ as well. 

Alternatively, 
when $\bbS = \bbL$ represents a normalized Laplacian \cite{EmergingFieldGSP}, the associated $\ccalS_{\mathrm{L}}$ is
\begin{align}\label{E:def_S_normalized_laplacian}
	\ccalS_{\rm L} \!\! := \! \{ \bbS \, | \, S_{ij}\! \in\! [-\!1, 0]   \,\, \text{for} \,\,& i\!\neq\! j,  \;\;\bbS\! \in \! \ccalM_{+}^N, \nonumber \\
	&S_{ii}\!=\!1 \,\, \forall \,\, i, \,\,\, \lambda_1 = 0 \} .
\end{align}
In $\ccalS_{\rm L}$ we impose that $\bbS$ is symmetric and positive semi-definite, its diagonal entries are $1$ and its off-diagonal entries are non-positive. Moreover, since $\bbS$ is a normalized Laplacian we know that the vector $\sqrt{\bbd}$ containing as entries the square roots of the node degrees is an eigenvector whose associated eigenvalue is zero, and this is incorporated into the last constraint. Notice that for this last constraint to be implementable we should be able to identify $\sqrt{\bbd}$ among all the spectral templates in $\bbV$. This can be done since $\sqrt{\bbd}$ is the only eigenvector whose entries have all the same sign \cite{biyikougu2007laplacian}. In the same way that fixing a scale discards the solution $\bbS = \mathbf{0}$ for adjacency matrices, the constraint $\lambda_1 = 0$ rules out the uninformative solution $\bbS = \bbI$ from the feasible set $\ccalS_{\rm L}$.

Naturally, the identification of other GSOs can be of interest as well, including for instance the combinatorial Laplacian $\bbL_c$ 
and the random walk Laplacian~\cite{chung1997spectral}. These can be accommodated in our proposed framework via minor modifications to the set $\ccalS$. For concreteness, we henceforth focus exclusively on adjacency and normalized Laplacian matrices.

%
\subsection{Additional sources for the spectral templates}\label{Ss:remark_templates} 

The central focus of this paper is to solve the problems in \eqref{eqn_zero_norm} and \eqref{E:general_problem} when eigenvectors $\bbv_k$ are estimated from a sample set $\ccalX$ (cf. Problem \ref{problem_data_driven}). Notwithstanding, the network topology inference problems in \eqref{eqn_zero_norm} and \eqref{E:general_problem} are applicable as long as eigenvectors or eigenvector estimates are available. Four examples are outlined next.

\myparagraphtc{GSO associated with orthogonal transformations}\label{ssec_AddSourcesGFT}
Expressing signals $\bbx$ in an alternative domain $\tbx$ by using an orthonormal transform $\tilde{\bbx} := \bbU^T\bbx$, such as Fourier, wavelets, or discrete-cosine, is a cornerstone operation in signal processing. If we make $\bbV=\bbU$ in \eqref{eqn_zero_norm} and \eqref{E:general_problem} we formulate the problem of identifying a graph shift $\bbS = \bbV \bbLambda \bbV^T = \bbU \bbLambda \bbU^T$ whose graph Fourier transform \cite{SandryMouraSPG_TSP14Freq} $\tilde{\bbx} := \bbV^T\bbx = \bbU^T\bbx$ is the given orthonormal transform of interest. This is important because it reveals the proximity structure between signal components that is implicitly assumed and exploited by the transform $\bbU$. 

\myparagraphtc{Design of graph filters}\label{sssec_linearoperators}
In addition to describing linear diffusion dynamics [cf.~\eqref{E:Filter_input_output_time}], graph filters represent linear transformations that can be implemented in a distributed manner~\cite{segarra2015reconstruction, EUSIPCO_our_interp_2015,ssamar_distfilters_allerton15}. In the context of distributed algorithms, consider implementing a prescribed linear network operator $\bbB$ using a graph filter $\bbH=\sum_{l=0}^{N-1}h_l\bbS^l$~\cite{segarra2015graphfilteringTSP15}. A necessary condition to accomplish this goal is that the eigenvectors of the shift $\bbS = \bbV \bbLambda \bbV^T$ and those of the linear transformation $\bbB = \bbV_\bbB \bbLambda_\bbB \bbV_\bbB^T$ must coincide; \cite[Prop. 1]{segarra2015graphfilteringTSP15}. Since $\bbV_\bbB$ can be obtained from the prescribed $\bbB$, the problems in \eqref{eqn_zero_norm} and \eqref{E:general_problem} can be solved using $\bbV=\bbV_\bbB$ as input. Problem \eqref{eqn_zero_norm}, for example, enable us to find the sparsest $\bbS$ which facilitates implementation of a given network operator $\bbB$ via distributed graph filtering. 

\myparagraphtc{Graph sparsification}\label{sssec_sparsify} Given a GSO $\bbT$, we can use our framework to obtain a different shift $\bbS$ with the same eigenvectors as $\bbT$, but with desirable properties encoded in $\ccalS$ and $f(\bbS,\bblambda)$ [cf.~\eqref{E:general_problem}]. If we set $f(\bbS,\bblambda)=\|\bbS\|_0$, this \emph{graph sparsification} problem can be addressed by solving \eqref{eqn_zero_norm} using as inputs the eigenvectors of $\bbT$. Note that, different from the setup in Problem \ref{problem_known_matrix}, the matrix $\bbT$ is not necessarily a covariance matrix.

\myparagraphtc{Network deconvolution}\label{sssec_network_deonvolution} The network deconvolution problem is the identification of an adjacency matrix $\bbS$ that encodes direct dependencies when given an adjacency $\bbT$ that includes indirect relationships. The problem is a generalization of channel deconvolution and can be solved by making $\bbS = \bbT \, (\bbI + \bbT)^{-1}$ \cite{FeiziNetworkDeconvolution}. This solution assumes a diffusion as in \eqref{eqn_diffusion} that results in a single-pole-single-zero graph filter. A more general approach is to assume that $\bbT$ can be written as a polynomial of $\bbS$ but be agnostic to the form of the filter. This leads to problem formulations \eqref{eqn_zero_norm} and \eqref{E:general_problem} with $\bbV$ given by the eigenvectors of $\bbT$. As in the graph sparsification problem and different from Problem \ref{problem_known_matrix}, the matrix $\bbT$ is not necessarily a covariance matrix.


\section{Topology inference from spectral templates}\label{S:ShiftInfFullEigen}


As discussed in the previous section, the goal is to find a graph shift $\bbS$ that is diagonalized by the given spectral templates $\bbV=[\bbv_1,\ldots,\bbv_N]$. 
In the absence of additional constraints the problem is ill-posed, so we further impose conditions on $\bbS$ via the set $\ccalS$ 
and search for the shift that minimizes a pre-specified cost $f$ [cf.~\eqref{E:general_problem}]. 

The structure of the feasible set in \eqref{E:general_problem} plays a critical role towards solving our network topology inference problem. The reason is twofold. First, notice that both sets $\ccalS_{\rm A}$ and $\ccalS_{\rm L}$ are convex. Hence,  convexity of problem \eqref{E:general_problem} depends exclusively on the choice of the objective $f(\bbS,\bblambda)$, a key property to facilitate the solution of \eqref{E:general_problem} in practice. If $f(\bbS,\bblambda)$ is chosen to be convex -- e.g., equal to $\|\bbS\|_p$ with $p\geq 1$ -- the overall optimization will be convex too. Second, the dimension of the feasible set is generally small. In fact, it can be shown that in a number of setups the feasible set reduces to a singleton, or otherwise to a low-dimensional subspace. This is important because even if the objective is non-convex, searching over a small space need not be necessarily difficult.  

We first investigate the size of the feasible set and provide conditions under which it reduces to a singleton thus rendering $f$ inconsequential to the optimization. Then, for the cases where there are multiple feasible solutions, we focus on the sparsity-promoting formulation, i.e., $f(\bbS,\bblambda)=\|\bbS\|_0$.  The resultant optimization is non-convex and in fact NP-hard, so we propose computationally-efficient convex relaxations which are provably tight under some technical conditions.

\subsection{Size of the feasibility set}\label{Ss:size_feasible_set}

The feasible set of problem \eqref{E:general_problem} for both $\ccalS_{\mathrm{L}}$ and $\ccalS_{\mathrm{A}}$ is in general small. To be more precise, some notation must be introduced. Define $\bbW  \! := \! \bbV  \odot \bbV \! \in \! \reals^{N^2 \! \times \! N}$, where $\odot$ denotes the Khatri-Rao product. Notice that from the definition of $\bbS$ we can write $\bbs := \mathrm{vec}(\bbS)$ as $\bbs = \bbW \bblambda$. Hence, each row of $\bbW$ represents the $N$ weighting coefficients that map $\bblambda$ to the corresponding entry of $\bbS$. Further, define the set $\ccalD$ containing the indices of $\bbs$ corresponding to the diagonal entries of $\bbS$ and select the 
corresponding rows of $\bbW$ to form $\bbW_\ccalD\!\in\!\reals^{N\times N}$. Also, define the matrix $\bbU := \bbV^{\mathbf{1}} \circ  \bbV^{\mathbf{1}}\in\! \reals^{N\times N}$, where $\circ$ denotes the elementwise product and $\bbV^{\mathbf{1}} := [\bbone, \bbv_2, \bbv_3, \ldots, \bbv_N ]$. Using these conventions, the following result holds.

\begin{myproposition}\label{P:Feasibility_singleton}
	Assume that \eqref{E:general_problem} is feasible, then it holds that:\\
	a) If $\ccalS = \ccalS_{\mathrm{A}}$, then $\mathrm{rank}({\bbW}_{\ccalD}) \leq N - 1$. Similarly, if $\ccalS = \ccalS_{\mathrm{L}}$, then $\mathrm{rank}(\bbU) \leq N - 1$.\\
	b) If $\mathrm{rank}({\bbW}_{\ccalD}) = N - 1$ when $\ccalS = \ccalS_{\mathrm{A}}$ or $\mathrm{rank}(\bbU) = N - 1$ when $\ccalS = \ccalS_{\mathrm{L}}$, then the feasible set of \eqref{E:general_problem} is a singleton. 
\end{myproposition}
\begin{myproof}
	We show statements a) and b) for the case $\ccalS = \ccalS_{\mathrm{A}}$. The proofs for $\ccalS = \ccalS_{\mathrm{L}}$ are analogous and thus omitted.
	The key of the proof is to note that we may write $\bbW_\ccalD \bblambda = \diag( \bbS ) = \mathbf{0}$ for all feasible $\bblambda$. Hence, feasibility implies that ${\bbW}_{\ccalD}$ is rank-deficient as stated in a).
	To show b), assume that $\mathrm{rank}({\bbW}_{\ccalD}) = N - 1$ so that $\bblambda$ in $\mathrm{null}({\bbW}_{\ccalD})$ is unique up to a scaling factor. However, since one of the conditions in $\ccalS_A$ forces the first row of $\bbS$ to sum up to $1$ [cf.~\eqref{E:SparseAdj_def_S}], this scaling ambiguity is resolved and the unique feasible $\bblambda$ (and hence $\bbS$) is obtained.
\end{myproof}

Proposition~\ref{P:Feasibility_singleton} offers sufficient conditions under which \eqref{E:general_problem} reduces to a feasibility problem. More specifically, when condition b) is met, the objective in \eqref{E:general_problem} is inconsequential since there exists only one feasible $\bbS$. For more general cases, however, the GSO that minimizes the particular cost function $f$ chosen is recovered. 
Among the potential cost functions, the sparsity-inducing $\ell_0$ norm $f(\bbS, \bblambda) = \| \bbS \|_0$ is non-convex, thus, challenging to solve in practice. Due to the widespread interest in identifying sparse graphs (e.g., of direct relationships among signal elements), we devote the ensuing subsection to study this latter case separately.

\subsection{Relaxation for the sparse formulation}\label{Ss:relaxation}

Many large-scale, real-world networks are sparse~\cite{kolaczyk2009book}, so it is often meaningful  to infer a sparse GSO where most of the entries in $\bbS$ are zero [cf.~\eqref{eqn_zero_norm}]. In practice, the usual approach to handle the non-convex $\ell_0$ (pseudo) norm objective in \eqref{eqn_zero_norm} is to relax it to an iteratively re-weighted $\ell_1$ norm. Specifically, with $p$ denoting an iteration index, we aim to solve a sequence $p=1,...,P$ of weighted $\ell_1$-norm minimization problems
\begin{align}\label{E:SparseAdj_wl11_obj}
	\bbS_{\omega}^* \! := \! \argmin_{ \{\bbS, \bblambda\}}\sum_{i,j} \!\omega_{ij}(p)|S_{ij}| \quad \text{s. to } \,\bbS=\sum_{k =1}^N \! \lambda_k\bbv_k \bbv_k^T, \,\bbS \!\in\! \ccalS,
\end{align}
with weights $\omega_{ij}(p):=\tau/\left(|S_{ij}(p-1)|+\delta\right)$, for appropriately chosen positive constants $\tau$ and $\delta$. Intuitively, the goal of the re-weighted scheme in \eqref{E:SparseAdj_wl11_obj} is that if $|S_{ij}(p-1)|$ is small, in the next iteration the penalization $\omega_{ij}(p)$ is large, promoting further shrinkage of $S_{ij}$ towards zero \cite{candes_l0_surrogate}.

Naturally, under condition b) in Proposition~\ref{P:Feasibility_singleton} the solutions $\bbS_0^*$ of \eqref{eqn_zero_norm} and $\bbS_{\omega}^*$ of \eqref{E:SparseAdj_wl11_obj} are guaranteed to coincide given that the feasible set is reduced to a singleton. Moreover, even when condition b) is not satisfied, there exist weights $\omega_{ij}$ that guarantee the equivalence of both solutions. 
To state this formally, define the set $\ccalJ$ containing the indices identifying the support of $\bbS^*_0$ and denote by $\ccalJ^c$ its complement. Whenever $\bbS_0^*$ is the unique solution to \eqref{eqn_zero_norm}, it is not hard to establish that by setting weights in \eqref{E:SparseAdj_wl11_obj} as $\omega_{ij}=1$ for $(i,j)\in\ccalJ^c$ and $\omega_{ij}=0$ otherwise, then $\bbS_{\omega}^*$ is unique and equal to $\bbS_0^*$. 

The upshot of this simple observation is that there exist optimal weights so that the sparsest solution $\bbS^*_0$ can be recovered by solving a convex optimization problem. This result confers validity to the re-weighted formulation in \eqref{E:SparseAdj_wl11_obj}, nonetheless, we can neither choose these weights without knowing $\bbS_0^*$ a priori nor there is a guarantee that the succession of weights $\omega_{ij}(p)$ converges to these optimal weights. Hence, we now focus on the derivation of theoretical guarantees for a {particular} set of weights that can be set a priori, namely, we consider the formulation in which each entry of the GSO is equally weighted. This boils down to solving the convex optimization problem
\begin{align} \label{E:SparseAdj_l01_obj}
	\bbS^*_1 \! := \! \argmin_{ \{\bbS, \bblambda\}} \,\| \bbS\|_{1} \quad \text{s. to } \bbS \! = \! \textstyle\sum_{k =1}^N \lambda_k\bbv_k \bbv_k^T, \,\,\, \bbS \in \ccalS.
\end{align}
Interestingly, under certain conditions we can ensure that the solution $\bbS^*_1$ to the relaxed problem \eqref{E:SparseAdj_l01_obj} coincides with $\bbS^*_0$. To be more specific, define $\bbs_0^*:= \mathrm{vec}(\bbS_0^*)$, denote by $\ccalD^c$ the complement of $\ccalD$ and partition $\ccalD^c$ into $\ccalK$ and $\ccalK^c$, with the former indicating the positions of the nonzero entries of $\bbs_{0\ccalD^c}^*:=(\bbs_0^*)_{\ccalD^c}$, where we recall that matrix \textit{calligraphic subscripts} select rows. Denoting by $^\dag$ the matrix pseudo-inverse, we define
\begin{equation}\label{E:def_matrix_M}
	\bbM := (\bbI - \bbW \bbW^\dag)_{\ccalD^c} \,\in\reals^{N^2-N\times N^2},
\end{equation}
i.e., the orthogonal projector onto the kernel of $\bbW^T$ constrained to the off-diagonal elements in $\ccalD^c$. With $\bbe_1$ denoting the first canonical basis vector, we construct the matrix 
\begin{equation}\label{E:def_matrix_R}
	\bbR := [ \bbM, \, \bbe_1 \otimes \bbone_{N-1}]\,\in\reals^{N^2-N\times N^2+1},
\end{equation}
by horizontally concatenating $\bbM$ and a column vector of size $|\ccalD^c|$ with ones in the first $N-1$ positions and zeros elsewhere. With this notation in place, the following recovery result holds.

\begin{mytheorem}\label{T:Recovery}
	Whenever $\ccalS = \ccalS_{\rm A}$ and assuming problem \eqref{E:SparseAdj_l01_obj} is feasible, $\bbS^*_1 = \bbS^*_0$ if the two following conditions are satisfied:\\
	A-1) $\rank(\bbR_{\ccalK}) = |\ccalK|$; and \\   
	A-2) There exists a constant $\delta > 0$ such that 
	\begin{equation}\label{E:condition_recovery_noiseless}
		{\psi_{\bbR}} := \| \bbI_{\ccalK^c}(\delta^{-2}\bbR \bbR^T+\bbI_{\ccalK^c}^T \bbI_{\ccalK^c})^{-1}\bbI_{\ccalK}^T \|_{M(\infty)}<1.
	\end{equation}
\end{mytheorem}
\begin{myproof}
	Recalling that $\bbs = \mathrm{vec}(\bbS)$, problem \eqref{E:SparseAdj_l01_obj} for the case where $\ccalS = \ccalS_{\rm A}$ can be reformulated as
	\begin{equation}\label{E:proof_noiseless_recovery_010}
		\min_{ \{\bbs, \bblambda\}} \;\; \|\bbs\|_1 \;\;\text{s. to } \;\;\bbs= \bbW \bblambda,\;\;\bbs_{\ccalD}=\bbzero,\;\;(\bbe_1 \otimes \bbone_N)^T \bbs = 1,
	\end{equation}
	where the last equality imposes that the first column of $\bbS$ must sum up to $1$ [cf.~\eqref{E:SparseAdj_def_S}]. Notice that the non-negativity constraint in $\ccalS_{\rm A}$ is ignored in \eqref{E:proof_noiseless_recovery_010}. However, if we show that \eqref{E:proof_noiseless_recovery_010} can recover the sparse solution $\bbs_0^*$, then the same solution would be recovered by the more constrained problem \eqref{E:SparseAdj_l01_obj}. Notice that we may solve for $\bblambda$ in closed form as $\bblambda^* = \bbW^{\dag} \bbs$. Consequently,  \eqref{E:proof_noiseless_recovery_010} becomes 
	\begin{equation}\label{E:proof_noiseless_recovery_020}
		\min_{\bbs} \|\bbs\|_1 \;\;\: \text{s. to } (\bbI - \bbW \bbW^\dag) \bbs = \bbzero,\;\bbs_{\ccalD}=\bbzero,\;(\bbe_1 \otimes \bbone_N)^T \bbs = 1.
	\end{equation}
	Leveraging the fact that $\bbI - \bbW \bbW^\dag$ is symmetric, the first equality in \eqref{E:proof_noiseless_recovery_020} can be rewritten as [cf.~\eqref{E:def_matrix_M}]
	\begin{equation}\label{E:proof_noiseless_recovery_030}
		(\bbI - \bbW \bbW^\dag)^T_\ccalD \bbs_{\ccalD} + \bbM^T \bbs_{\ccalD^c} = \bbzero,
	\end{equation}
	and the second equality in \eqref{E:proof_noiseless_recovery_020} forces the first term of \eqref{E:proof_noiseless_recovery_030} to be zero. With these considerations, we may restate \eqref{E:proof_noiseless_recovery_020} as 
	\begin{equation}\label{E:proof_noiseless_recovery_040}
		\min_{ \bbs_{\ccalD^c}} \;\; \|\bbs_{\ccalD^c}\|_1 \;\;\text{s. to } \;\; \bbR^T \bbs_{\ccalD^c} = \bbb,
	\end{equation}
	where $\bbb$ is a binary vector of length $N^2+1$ with all its entries equal to $0$ except for the last one that is a $1$.
	Problem \eqref{E:proof_noiseless_recovery_040} takes the form of classical basis pursuit~\cite{ChenDonohoBP}. Notice that the system of linear equations in \eqref{E:proof_noiseless_recovery_040} is overdetermined since $\bbR^T \in \reals^{N^2+1 \times | \ccalD^c|}$, however, feasibility of \eqref{E:proof_noiseless_recovery_010} guarantees that the mentioned system of equations is compatible. The following two conditions are required for the solution of \eqref{E:proof_noiseless_recovery_040} to coincide with the sparse solution ${\bbs_0^*}_{\ccalD^c}$ (cf.~\cite{zhang2013one}): 
	\begin{itemize}
		\item[a)] $\mathrm{ker}(\bbI_{\ccalK^c}) \cap \mathrm{ker}(\bbR^T) = \{ \mathbf{0} \}$; and
		\item[b)] There exists a vector $\bby \in \reals^{|\ccalD^c|}$ such that $\bby \in \mathrm{Im}(\bbR)$, $\bby_\ccalK = \mathrm{sign}(({\bbs_0^*}_{\ccalD^c})_\ccalK)$, and $\| \bby_{\ccalK^c}\|_\infty < 1$.
	\end{itemize}
	
	\noindent The remainder of the proof is devoted to showing that if conditions \emph{A-1)} and \emph{A-2)} in the statement of the theorem hold true, then a) and b) are satisfied.
	
	To see that \emph{A-1)} implies a) notice that the nullspace of $\bbI_{\ccalK^c}$ is spanned by the columns of $\bbI^T_{\ccalK}$. Hence, for a) to hold we need the $|\ccalK|$ columns of $\bbR^T$ in positions $\ccalK$ to form a full column rank matrix. In condition \emph{A-1)} we require $\bbR_\ccalK$ to be full row rank, which is an equivalent property.
	
	The next step is to show that condition \emph{A-2)} implies b). For this, consider the following $\ell_2$-norm minimization problem
	\begin{align}\label{E:l_2_minimization_dual_certificate}
		\min_{ \{\bby, \bbz\}} \; \delta^2\|\bbz\|_2^2 + \|\bby\|_2^2  \;\;\text{s. to } \; \bby= \bbR\bbz,\;\; \bby_\ccalK = \mathrm{sign}(({\bbs_0^*}_{\ccalD^c})_\ccalK), 
	\end{align}
	where $\delta$ is a positive tuning constant. The inclusion of the term $\delta^2\|\bbz\|_2^2$ in the objective guarantees the existence of a closed-form expression for the optimal solution, while preventing numerical instability when solving the optimization. We will show that the solution $\bby^*$ to problem \eqref{E:l_2_minimization_dual_certificate} satisfies the requirements imposed in condition b). The two constraints in \eqref{E:l_2_minimization_dual_certificate} enforce the fulfillment of the first two requirements in b), hence, we are left to show that $\| \bby^*_{\ccalK^c}\|_\infty < 1$.
	Since the values of $\bby_{\ccalK}$ are fixed, the constraint $\bby= \bbR \bbz$ can be rewritten as $\bbI_{\ccalK}^T \mathrm{sign}(({\bbs_0^*}_{\ccalD^c})_\ccalK) = -\bbI_{\ccalK^c}^T \bby_{\ccalK^c} +  \bbR \delta^{-1} \delta\bbz$. Then, by defining the vector $\bbt := [\delta\bbz^T, -\bby_{\ccalK^c}^T]^T$ and the matrix $\bbPhi:= [\delta^{-1} \bbR^T, \bbI_{\ccalK^c}]$, \eqref{E:l_2_minimization_dual_certificate} can be rewritten as
	\begin{align}\label{E:l_2_minimization_dual_certificate_rewritten}
		\min_{\bbt} \;\; \|\bbt\|_2^2  \;\;\text{s. to } \; \bbI_{\ccalK}^T \mathrm{sign}(({\bbs_0^*}_{\ccalD^c})_\ccalK) = \bbPhi^T \bbt.
	\end{align}
	The minimum-norm solution to \eqref{E:l_2_minimization_dual_certificate_rewritten} is given by $\bbt^*=(\bbPhi^T)^\dag \bbI_{\ccalK}^T \mathrm{sign}(({\bbs_0^*}_{\ccalD^c})_\ccalK)$ from where it follows that
	\begin{align}\label{E:dual_certificate_solution_norm2_delta}
		\bby^*_{\ccalK^c}  \!=\!  - \bbI_{\ccalK^c}(\delta^{-2}\bbR \bbR^T+\bbI_{\ccalK^c}^T \bbI_{\ccalK^c})^{-1}\bbI_{\ccalK}^T \, \mathrm{sign}(({\bbs_0^*}_{\ccalD^c})_\ccalK).
	\end{align}
	Condition a) guarantees the existence of the inverse in \eqref{E:dual_certificate_solution_norm2_delta}. Since $\| \mathrm{sign}(({\bbs_0^*}_{\ccalD^c})_\ccalK) \|_{\infty} \! = \! 1$, we may bound the $\ell_\infty$ norm of $\bby^*_{\ccalK^c}$ as $\| \bby^*_{\ccalK^c} \|_{\infty} \leq \| \bbI_{\ccalK^c}(\delta^{-2}\bbR \bbR^T+\bbI_{\ccalK^c}^T \bbI_{\ccalK^c})^{-1}\bbI_{\ccalK}^T \|_{M(\infty)} = \psi_{\bbR}$.
	Hence, condition \emph{A-2)} in the theorem guarantees $\| \bby^*_{\ccalK^c} \|_{\infty}<1$ as wanted, concluding the proof.
\end{myproof}


Theorem~\ref{T:Recovery} offers \textit{sufficient} conditions under which the relaxation \eqref{E:SparseAdj_l01_obj} guarantees sparse recovery for adjacency matrices. Simulations in Section~\ref{S:Simulations} reveal that the bound imposed on $\psi_{\bbR}$ is tight by providing examples where $\psi_{\bbR}$ is equal to 1 and for which recovery fails. In Theorem~\ref{T:Recovery}, condition \emph{A-1)} ensures that the solution to \eqref{E:SparseAdj_l01_obj} is unique, a necessary requirement to guarantee sparse recovery. Condition \emph{A-2)} is derived from the construction of a dual certificate specially designed to ensure that the unique solution to \eqref{E:SparseAdj_l01_obj} also has minimum $\ell_0$ norm~\cite{zhang2013one}.

Recall that the $\ell_\infty$ norm in \eqref{E:condition_recovery_noiseless} is the maximum $\ell_1$ norm across the rows of the argument matrix, which has $|\ccalK_c|$ rows each containing $|\ccalK|$ elements. It is thus expected that sparser graphs (small $|\ccalK|$) might have smaller values of $\psi_{\bbR}$. Furthermore, to have an intuitive understanding of $\psi_{\bbR}$ it is helpful to see that condition \emph{A-2)} is always satisfied whenever $\bbR \bbR^T$ is non-singular. More specifically, for small values of $\delta$ we have that ${\psi_{\bbR}}  \approx \delta^2 \| \bbI_{\ccalK^c}(\bbR \bbR^T)^{-1}\bbI_{\ccalK}^T \|_{M(\infty)}$, which can be made arbitrarily small and, in particular, strictly smaller than 1. Matrix $\bbR \bbR^T$ can be shown to be invertible whenever $\mathrm{rank}({\bbW}_{\ccalD}) = N - 1$ (cf.~Proposition \ref{P:Feasibility_singleton}). Thus, in the extreme case where the feasible set is a singleton, Theorem~\ref{T:Recovery} guarantees recovery, as expected. A more general characterization of the ensembles of random graphs that tend to satisfy \eqref{E:condition_recovery_noiseless} with high probability is of interest, but left as future research.

The recovery result of Theorem~\ref{T:Recovery} can be replicated for the case where the shift of interest is a normalized Laplacian, i.e., when $\ccalS = \ccalS_{\rm L}$. To state this formally, if we define $\bbQ := (\bbI - \tilde{\bbU} \tilde{\bbU}^\dag)_{\ccalD^c}$, where $\tilde{\bbU} := \tilde{\bbV} \odot  \tilde{\bbV}$ for $ \tilde{\bbV} := [\bbv_2, \bbv_3, \ldots, \bbv_N]$ the following result holds. The proof follows the same steps as those in Theorem~\ref{T:Recovery} and, thus, is omitted.

\begin{mytheorem}\label{T:Recovery_Laplacian}
	Whenever $\ccalS = \ccalS_{\rm L}$ and assuming problem \eqref{E:SparseAdj_l01_obj} is feasible, $\bbS^*_1 = \bbS^*_0$ if the two following conditions are satisfied:\\
	L-1) $\rank(\bbQ_\ccalK) = | \ccalK |$; and \\
	L-2) There exists a constant $\delta > 0$ such that
	\begin{equation}\label{E:condition_recovery_noiseless_laplacian}
		{\psi_{\bbQ}} := \| \bbI_{\ccalK^c}(\delta^{-2}\bbQ \bbQ^T+\bbI_{\ccalK^c}^T \bbI_{\ccalK^c})^{-1}\bbI_{\ccalK}^T \|_{M(\infty)}<1.
	\end{equation}
\end{mytheorem}


\section{Imperfect spectral templates}\label{S:ShiftInfSubsetEigen}


Whenever the number of observed graph signals is limited or the observations are noisy, assuming perfect knowledge of the spectral templates $\bbV$ may be unrealistic. This section broadens the scope of the network inference problems dealt with so far, to accommodate imperfect spectral templates that can either be noisy or incomplete. Specifically, we investigate pragmatic scenarios where: i) only an approximate version of $\bbV$ can be obtained (e.g., from the eigenvectors of a \emph{sample} covariance matrix); and ii) where only a subset of $\bbV$ is available (e.g., when the observed signals are bandlimited and one can only estimate the non-zero frequencies that are present).

\subsection{Noisy spectral templates} \label{Ss:ShiftRecImperfEig}


We first address the case where knowledge of an approximate version of the spectral templates $\hat{\bbV} = [\hat{\bbv}_1,\ldots,\hat{\bbv}_N]$ is available.
The question here is how to update the general formulation in \eqref{E:general_problem} to account for the discrepancies between the estimated spectral templates $\hat{\bbV}$ and the actual eigenvectors of $\bbS$. An instructive reformulation is to include $\bbV= [\bbv_1, \ldots, \bbv_N]$ as decision variables and formulate the following problem
\begin{align}\label{E:SparseAdj_l1_obj_noisy_vectors}
	&\min_{\{\bbS, \bblambda, \bbV\}} \;f( \bbS,\bblambda) \quad \\
	&\text{s. to } \,\, \bbS = \textstyle\sum_{k =1}^N \lambda_k\bbv_k \bbv_k^T, \,\,\, \bbS \in \ccalS, \,\,\, d(\bbv_k, \hat{\bbv}_k) \leq \epsilon_k \,\,\forall \, k, \nonumber
\end{align}
where $d(\cdot, \cdot)$ is a \textit{convex} vector distance function, such as the $\ell_p$ norm of the vector difference for $p \! \geq \! 1$. The idea in \eqref{E:SparseAdj_l1_obj_noisy_vectors} is to find a sparse $\bbS$ that satisfies the desired properties in $\ccalS$, while its eigenvectors $\bbv_k$ are each of them close to the observed ones $\hat{\bbv}_k$. The value of $\epsilon_k$ must be chosen based on a priori information on the imperfections, such as the number of signals used to estimate the sample covariance, or the statistics of the observation noise. While conceptually simple, problem \eqref{E:SparseAdj_l1_obj_noisy_vectors} is more challenging than its noiseless counterpart, since the first constraint is non-convex given that both $\lambda_k$ and $\bbv_k$ are optimization variables. 

A more tractable alternative is to form $\bbS':=\sum_{k =1}^{N} \lambda_k \hat{\bbv}_k \hat{\bbv}_k^T$ and search for a shift $\bbS$ that possesses the desired properties while being close to $\bbS'$. Formally, one can solve
\begin{align}\label{E:SparseAdj_l1_obj_noisy_matrix}
	\hat{\bbS}^*&:=\argmin_{\{\bbS, \bblambda, \bbS'\}} \; f( \bbS,\bblambda) \,\,\,\, \\
	&\text{s. to } \,\, \bbS' = \textstyle\sum_{k =1}^N \lambda_k \hat{\bbv}_k \hat{\bbv}_k^T, \,\,\,\, \bbS \in \ccalS, \,\,\,d(\bbS, \bbS') \leq \epsilon, \nonumber
\end{align}
where $d(\cdot, \cdot)$ is a \textit{convex} matrix distance whose form depends on the particular application. E.g., if $\|\bbS-\bbS'\|_\mathrm{F}$ is chosen, the focus is more on the similarities across the entries of the shifts, while $\|\bbS-\bbS'\|_{M(2)}$ focuses on their spectrum. Additional conic constraints of the form $\|(\bbS-\bbS')\hbv_k\|_2\leq \lambda_k\epsilon_k$ enforcing that particular eigenvectors are well approximated can also be incorporated. From an application point of view, the formulation in \eqref{E:SparseAdj_l1_obj_noisy_matrix} is also relevant to setups where the templates $\hbV$ are not necessarily noisy but the goal is to enlarge the set of feasible GSOs. This can be of interest if, for example, finding an $\bbS$ that is both sparse and with the exact templates collected in $\hbV$ is impossible (cf. Section \ref{Ss:remark_templates}).

The difficulty in solving \eqref{E:SparseAdj_l1_obj_noisy_matrix} is determined by $f$ and $\ccalS$. Hence, the challenges and approaches are similar to those in Section \ref{S:ShiftInfFullEigen}. For the particular case of sparse shifts, the $\ell_1$ norm relaxation of \eqref{E:SparseAdj_l1_obj_noisy_matrix} yields [cf.~\eqref{E:SparseAdj_l01_obj}]
\begin{align}\label{E:SparseAdj_l1_obj_noisy_matrix_v2}
	\hat{\bbS}^*_1&:=\argmin_{\{\bbS, \bblambda, \bbS'\}} \; \|\bbS\|_1 \,\,\,\, \\
	&\text{s. to } \,\, \bbS' = \textstyle\sum_{k =1}^N \lambda_k \hat{\bbv}_k \hat{\bbv}_k^T, \,\,\,\, \bbS \in \ccalS, \,\,\,d(\bbS, \bbS') \leq \epsilon, \nonumber
\end{align}
where iteratively re-weighted schemes are also possible. Moreover, further uncertainties can be introduced in the definition of the feasible set $\ccalS$, e.g. in the scale of the admissible graphs for the case of $\ccalS = \ccalS_{\rm A}$ (cf.~Proposition~\ref{P:noisy_recovery} and \eqref{E:proof_noisy_recovery_005} for additional details). 

When the interest is in recovering a normalized Laplacian [cf.~\eqref{E:def_S_normalized_laplacian}], a possible implementation is to enforce the constraint $\lambda_1 = 0$ \emph{talis qualis} on \eqref{E:SparseAdj_l1_obj_noisy_matrix_v2} entailing that one of the eigenvalues of $\bbS'$ (and not $\bbS$) is equal to zero. However, the smallest eigenvalue of $\bbS$ must be close to zero due to the constraint on the distance between $\bbS$ and $\bbS'$. Alternatively, the objective can be augmented by also considering the nuclear norm $\| \bbS \|_*$ to further promote rank-deficiency on $\bbS$.

To assess the effect of the noise in recovering the sparsest $\bbS$, we define matrices $\hbW$, $\hat{\bbR}$ and $\hat{\bbQ}$ which are counterparts of $\bbW$, $\bbR$ and $\bbQ$ defined prior to Theorem~\ref{T:Recovery}, but based on the noisy templates $\hat{\bbV}$ instead of $\bbV$. Further, we drop the non-negativity constraint in $\ccalS_{\rm A}$ -- to obtain $\tilde{\ccalS}_{\rm A}$ -- and incorporate the scale ambiguity by augmenting $d(\bbS, \bbS')$ as $\tilde{d}(\bbS, \bbS') = (d(\bbS, \bbS')^2 + (\textstyle\sum_j S_{j1} - 1)^2)^{1/2}$. With this notation, the following result on robust recovery of network topologies holds.

%
\begin{myproposition}\label{P:noisy_recovery}
	When $d(\bbS, {\bbS}') = \| \bbS - {\bbS}' \|_\mathrm{F}$, and assuming that there exists at least one ${\bbS}'$ such that $\tilde{d}(\bbS_0^*, {\bbS}')\leq \epsilon$, the solution $\hbs_1^* := \mathrm{vec}(\hbS_1^*)$ to \eqref{E:SparseAdj_l1_obj_noisy_matrix_v2} for $\ccalS = \tilde{\ccalS}_{\rm A}$ with scale ambiguity satisfies
	\begin{equation}\label{E:recov_noise}
		\| \hbs_1^* - \bbs_0^* \|_1 \leq C \epsilon, \quad \text{with} \,\,\, C = 2 C_1 + 2 C_2 C_3,
	\end{equation}
	if the same conditions stated in Theorem~\ref{T:Recovery} hold but for $\hat{\bbR}$ instead of ${\bbR}$.
	Constants $C_1$, $C_2$, and $C_3$ are given by
	\begin{equation}\label{E:recov_noise_constants}
		C_1 \!=\! \frac{\sqrt{| \ccalK |}}{\sigma_{\min}(\hat{\bbR}^T_\ccalK)}, \,\, C_2 \!=\! \frac{1 + \| \hat{\bbR}^T \|_{M(2)} C_1}{1 - \psi_{\hat{\bbR}}}, \,\, C_3 \!=\! \| \hat{\bbR}^{\dag} \|_{M(2)} N,
	\end{equation}
	where $\sigma_{\min}( \cdot)$ denotes the minimum singular value of the argument matrix.
	An analogous result can be derived for the case $\ccalS = \tilde{\ccalS}_\mathrm{L}$ (where the non-positivity constraint is dropped) whenever $\hat{\bbQ}$ satisfies the conditions in Theorem~\ref{T:Recovery_Laplacian}.
\end{myproposition}
\begin{myproof}
	We reformulate \eqref{E:SparseAdj_l1_obj_noisy_matrix_v2} in vector form for the case $\ccalS = \tilde{\ccalS}_{\rm A}$ with scale ambiguity to obtain
	\begin{align}\label{E:proof_noisy_recovery_005}
		\min_{ \{\bbs, \bblambda, \bbs'\}} \;\; \|\bbs\|_1 \;\;\text{s. to }& \;\;\bbs'= \hbW \bblambda,\;\;\bbs_{\ccalD}=\bbzero, \\
		&\|\bbs - \bbs'\|_2^2 + ((\bbe_1 \otimes \bbone_N)^T \bbs - 1)^2 \leq \epsilon^2. \nonumber
	\end{align}
	Substituting the first equality constraint in \eqref{E:proof_noisy_recovery_005} into the inequality constraint, then solving for $\bblambda$ as $\bblambda^* = \hbW^\dag \bbs$, and finally using the second equality constraint to reduce the optimization variables to $\bbs_{\ccalD^c}$, we may restate \eqref{E:proof_noisy_recovery_005} as [cf.~\eqref{E:proof_noiseless_recovery_040}]
	\begin{equation}\label{E:proof_noisy_recovery_010}
		\min_{ \bbs_{\ccalD^c}} \;\; \|\bbs_{\ccalD^c}\|_1 \quad\; \text{s. to } \; \| \hat{\bbR}^T \bbs_{\ccalD^c} - \bbb \|_2 \leq \epsilon,
	\end{equation}
	where $\bbb$ is, as in the proof of Theorem~\ref{T:Recovery}, a binary vector with all its entries equal to 0 except for the last one that is equal to 1. Notice that \eqref{E:proof_noisy_recovery_010} takes the form of a basis pursuit problem with noisy observations~\cite{ChenDonohoBP}. Expressions \eqref{E:recov_noise} and \eqref{E:recov_noise_constants} can be derived by applying the second claim in \cite[Theorem~2]{zhang2013one} to problem \eqref{E:proof_noisy_recovery_010}. In order to do so, a few factors must be taken into consideration. First, since $\hat{\bbR}^T$ is not full row rank (since it is a tall matrix), constant $C_3$ depends on the $\ell_2$ norm of $\hat{\bbR}^{\dag}$. Moreover, in order to make constants $C_1$, $C_2$, and $C_3$ independent of the dual certificate $\bby \in \reals^{|\ccalD^c|}$ -- see condition b) within the proof of Theorem~\ref{T:Recovery} -- we have used that $\| \bby \|_2 \leq N $ and $\| \bby_{\ccalK^c} \|_\infty \leq \psi_{\hat{\bbR}}$, where the first one follows from the fact that the magnitude of every element in $\bby$ is at most $1$ and the second one was shown after \eqref{E:dual_certificate_solution_norm2_delta}.
	
	A similar procedure can be used to show the result pertaining the case where $\ccalS \!= \! \tilde{\ccalS}_{\rm L}$.
\end{myproof}

When given noisy versions $\hat{\bbV}$ of the spectral templates of our target GSO, Proposition~\ref{P:noisy_recovery} quantifies the effect that the noise has on the recovery. More precisely, the recovered shift is guaranteed to be at a maximum distance from the desired shift bounded by the tolerance $\epsilon$ times a constant, which depends on $\hbR$ and the support $\ccalK$. In particular, this implies that as the number of observed signals increases we recover the true graph shift as stated in the following remark.

\begin{remark}[Consistent estimator]\label{R:consistent_estimator}\normalfont
	As the number of observed signals increases the sample covariance $\hbC_x$ tends to the covariance $\bbC_x$ and, for the cases where the latter has no repeated eigenvalues, the noisy eigenvectors $\hbV$ tend to the eigenvectors $\bbV$ of the desired shift; see, e.g., \cite[Theo. 3.3.7]{Ortega90}. In particular, with better estimates $\hat{\bbV}$ the tolerance $\epsilon$ in \eqref{E:SparseAdj_l1_obj_noisy_matrix_v2} needed to guarantee feasibility can be made smaller, entailing a smaller discrepancy between the recovered $\bbS_1^*$ and the sparsest shift $\bbS_0^*$. In the limit when $\hat{\bbV} = \bbV$ and under no additional uncertainties, the tolerance $\epsilon$ can be made zero and \eqref{E:recov_noise} guarantees perfect recovery under conditions \emph{A-1)} and \emph{A-2)} in Theorem~\ref{T:Recovery} or \emph{L-1)} and \emph{L-2)} in Theorem~\ref{T:Recovery_Laplacian}.
\end{remark}

\subsection{Incomplete spectral templates} \label{Ss:ShiftRecIncompEig}

Thus far we have assumed that the entire set of eigenvectors $\bbV= [\bbv_1, \ldots, \bbv_N]$ is known, either perfectly or corrupted by noise. However, it is conceivable that in a number of scenarios only some of the eigenvectors (say $K$ out of $N$) are available. This would be the case when e.g., $\bbV$ is found as the eigenbasis of $\bbC_x$ and the given signal ensemble is bandlimited. More generally, whenever $\bbC_x$ contains repeated eigenvalues there is a rotation ambiguity in the definition of the associated eigenvectors. Hence, in this case, we keep the eigenvectors that can be unambiguously characterized and, for the eigenvectors with repeated eigenvalues, we include the rotation ambiguity as an additional constraint in our optimization problem.

Formally, assume that the $K$ first eigenvectors $\bbV_K=[\bbv_1,...,\bbv_K]$ are those which are known. Then, the network topology inference problem with incomplete spectral templates can be formulated as [cf.~\eqref{E:SparseAdj_l01_obj}] 
\begin{align}\label{E:SparseAdj_l00_onlysomeeig}
	\bar{\bbS}^*_1& := \argmin_{ \{\bbS, \bbS_{\bar{K}}, \bblambda\}} \;\; \| \bbS \|_1 \\
	& \text{s. to } \;\; \bbS= \bbS_{\bar{K}} + {\textstyle\sum_{k=1}^{K}} \lambda_k \bbv_k \bbv_k^T, \;\; \bbS\in\ccalS, \;\; \bbS_{\bar{K}}\bbV_K=\bb0, \nonumber
\end{align}
where we already particularized the objective to the $\ell_1$ convex relaxation.
The formulation in \eqref{E:SparseAdj_l00_onlysomeeig} enforces $\bbS$ to be partially diagonalized by the known spectral templates $\bbV_K$, while its remaining component $\bbS_{\bar{K}}$ is forced to belong to the orthogonal complement of $\text{range}(\bbV_K)$. Notice that, as a consequence, the rank of $\bbS_{\bar{K}}$ is at most $N-K$. As in the previous cases, $\ccalS$ incorporates a priori information about the GSO. Notice that the constraint in $\ccalS$ imposing symmetry on $\bbS$ combined with the first constraint in \eqref{E:SparseAdj_l00_onlysomeeig} automatically enforce symmetry on $\bbS_{\bar{K}}$, as wanted.
An advantage of using only partial information of the eigenbasis as opposed to the whole $\bbV$ is that the set of feasible solutions in \eqref{E:SparseAdj_l00_onlysomeeig} is larger than that in \eqref{E:SparseAdj_l01_obj}. This is particularly important when the  templates do not come from a preexisting shift but, rather, one has the freedom to choose $\bbS$ provided it satisfies certain spectral properties. A practical example is the selection of the topology of a sensor network aimed at implementing estimation tasks such as consensus averaging, which can be oftentimes written as rank-one transformations of the sensor observations (cf. Section \ref{Ss:remark_templates} and~\cite{segarra2015graphfilteringTSP15}).

Theoretical guarantees of recovery analogous to those presented in Section~\ref{Ss:relaxation} can be derived for \eqref{E:SparseAdj_l00_onlysomeeig}. To formally state these, the following notation must be introduced. Define $\bbW_K := \bbV_K \odot \bbV_K$ and $\bbUpsilon := [\bbI_{N^2}, \bbzero_{N^2 \times N^2}]$. Also, define matrices $\bbB^{(i,j)} \in \reals^{N \times N}$ for $i < j$ such that $B^{(i,j)}_{ij} = 1$, $B^{(i,j)}_{ji} = -1$, and all other entries are zero. Based on this, we denote by $\bbB \in \reals^{{N \choose 2} \times N^2}$ a matrix whose rows are the vectorized forms of $\bbB^{(i,j)}$ for all $i,j \in \{1, 2, \ldots, N\}$ where $i < j$. In this way, $\bbB \bbs = \bbzero$ when $\bbs$ is the vectorized form of a symmetric matrix. Further, we define the following matrices 
\begin{equation}\label{E:def_P_matrices}
	\bbP_1 \!:=\! 
	\begin{bmatrix}
		\bbI - \bbW_K \bbW_K^\dag \\
		\bbI_\ccalD \\
		\bbB \\
		\bbzero_{NK \times N^2} \\
		(\bbe_1 \otimes \bbone_N)^T 
	\end{bmatrix}^T, \quad
	\bbP_2 \!:=\! 
	\begin{bmatrix}
		\bbW_K \bbW_K^\dag - \bbI \\
		\bbzero_{N \times N^2} \\
		\bbzero_{{N \choose 2} \times N^2} \\
		\bbI \otimes V_K^T \\
		\bbzero_{1 \times N^2}
	\end{bmatrix}^T,
\end{equation}
and $\bbP := [\bbP_1^T , \bbP_2^T]^T$. With this notation in place, and denoting by $\ccalJ$ the support of $\bbs_0^* = \mathrm{vec}(\bbS^*_0)$, the following result holds.

\begin{mytheorem}\label{T:Recovery_incomplete}
	Whenever $\ccalS = \ccalS_{\rm A}$ and assuming problem \eqref{E:SparseAdj_l00_onlysomeeig} is feasible, $\bar{\bbS}^*_1 = \bbS^*_0$ if the two following conditions are satisfied:\\
	A-1) $\rank([{\bbP_1}_\ccalJ^T, \bbP_2^T]) = |\ccalJ| + N^2$; and \\   
	A-2) There exists a constant $\delta > 0$ such that 
	\begin{equation}\label{E:condition_recovery_incomplete}
		\eta_{\bbP} := \| \bbUpsilon_{\ccalJ^c}(\delta^{-2} \bbP \bbP^T + \bbUpsilon_{\ccalJ^c}^T \bbUpsilon_{\ccalJ^c})^{-1}\bbUpsilon_{\ccalJ}^T \|_{M(\infty)}<1.
	\end{equation}
\end{mytheorem}
\begin{myproof}
	With $\bbs = \mathrm{vec}(\bbS)$ and $\bbs_{\bar{K}} = \mathrm{vec}(\bbS_{\bar{K}})$, we reformulate \eqref{E:SparseAdj_l00_onlysomeeig} for $\ccalS = \ccalS_{\rm A}$ as
	\begin{align}\label{E:proof_incomplete_recovery_010}
		\min_{ \{\bbs, \bbs_{\bar{K}}, \bblambda\}} & \;\; \| \bbs \|_1 \\
		\text{s. to } \;\; &\bbs= \bbs_{\bar{K}} + \bbW_K \bblambda, \; \bbs_{\ccalD} = \bbzero, \; \bbB \bbs = \bbzero,  \nonumber \\
		& (\bbe_1 \otimes \bbone_N)^T \bbs= 1,  \; (\bbI \otimes \bbV_K^T)\bbs_{\bar{K}} = \bbzero. \nonumber
	\end{align}
	The first and last constraints in \eqref{E:proof_incomplete_recovery_010} correspond to the first and last constraints in \eqref{E:SparseAdj_l00_onlysomeeig} written in vector form. The second constraint in \eqref{E:proof_incomplete_recovery_010} imposes that $\bbS$ has no self-loops, the third one imposes symmetry on $\bbS$, and the fourth one normalizes the first column of $\bbS$ to sum up to 1 [cf.~\eqref{E:SparseAdj_def_S}]. Notice that the non-negativity constraint in $\ccalS_{\rm A}$ is ignored in \eqref{E:proof_incomplete_recovery_010}; however, if we show that \eqref{E:proof_incomplete_recovery_010} can recover the sparse solution $\bbs_0^*$, then the same solution would be recovered by the more constrained problem \eqref{E:SparseAdj_l00_onlysomeeig}. Using the first constraint to solve for $\bblambda$, we obtain $\bblambda = \bbW_K^\dag (\bbs - \bbs_{\bar{K}})$. Moreover, defining the concatenated variable $\bbt:= [ \bbs^T, \bbs_{\bar{K}}^T]^T$, it follows from the definitions of $\bbUpsilon$ and $\bbP$ that \eqref{E:proof_incomplete_recovery_010} can be reformulated as
	\begin{align}\label{E:proof_incomplete_recovery_020}
		\min_{ \bbt }  \;\; \| \bbUpsilon \bbt \|_1 \quad \; \text{s. to } \;  \bbP^T \bbt = \bbb,
	\end{align}
	where $\bbb$ is a vector with every entry equal to $0$ except for the last one which is equal to $1$. We utilize existing results on $\ell_1$-analysis \cite{zhang2013one} to state that the solution to \eqref{E:proof_incomplete_recovery_020} coincides with the sparsest solution if:
	\begin{itemize}
		\item[a)] $\mathrm{ker}(\bbUpsilon_{\ccalJ^c}) \cap \mathrm{ker}(\bbP^T) = \{ \mathbf{0} \}$; and
		\item[b)] There exists a vector $\bby \in \reals^{N^2}$ such that $\bbUpsilon^T \bby \in \mathrm{Im}(\bbP)$, $\bby_\ccalJ = \mathrm{sign}({\bbs_0^*}_{\ccalJ})$, and $\|\bby_{\ccalJ^c}\|_\infty < 1$.
	\end{itemize}

	\begin{figure*}
		\centering
		\input{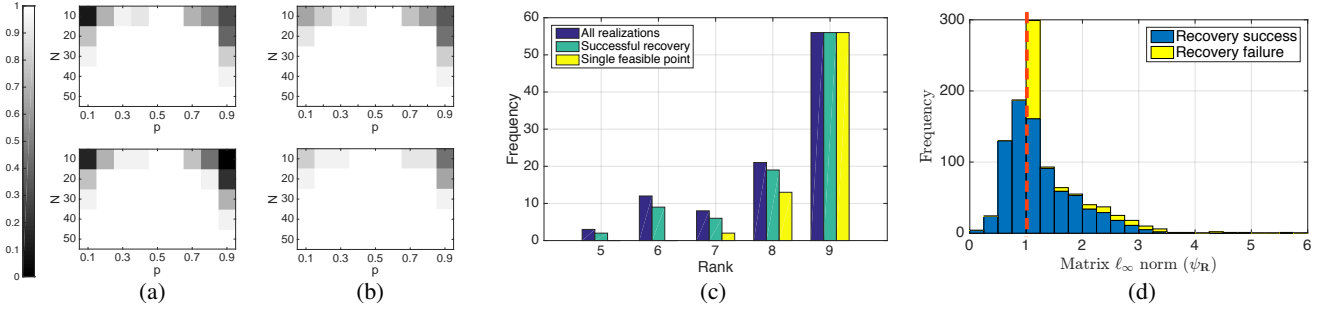}	
		\vspace{-0.15in}
		\caption{(a) Proportion of topology inference problems with a unique feasible point for Erd\H{o}s-R\'enyi graphs as a function of $N$ and $p$ for adjacency (top) and normalized Laplacian (bottom) matrices. (b) Recovery rate for the same set of graphs in (a) when implementing the iteratively re-weighted approach in \eqref{E:SparseAdj_wl11_obj}. (c) Histogram of the rank of matrix $\bbU$ for $N\!=\!10$ and $p\!=\!0.2$. (d) Experimental validation of Theorem~\ref{T:Recovery}. Whenever $\psi_{\bbR} < 1$ perfect recovery is achieved.}
		\vspace{-0.1in}
		\label{F:num_exp} 
	\end{figure*}
	
	\noindent As was the case for Theorem~\ref{T:Recovery}, the proof now reduces to showing that conditions \emph{A-1)} and \emph{A-2)} in the statement of the theorem imply the above conditions a) and b).
	
	From the specific form of $\bbUpsilon$, the kernel of $\bbUpsilon_{\ccalJ^c}$ is a space of dimension $|\ccalJ| + N^2$ spanned by the set of canonical basis vectors $\bbe_i$ of length $2 N^2$ where $i \in \ccalJ \cup \{N^2+1, N^2+2, \ldots, 2 N^2\}$. Thus for a) to hold we need the matrix formed by the columns of $\bbP^T$ indexed by $\ccalJ \cup \{N^2+1, \ldots, 2 N^2\}$ to be full column rank, as stated in condition \emph{A-1)}.
	
	Finally, the procedure to show that \emph{A-2)} implies b) follows the same steps as those detailed in the proof of Theorem~\ref{T:Recovery} -- from \eqref{E:l_2_minimization_dual_certificate} onwards -- and, thus, is omitted here. 
\end{myproof}

Theorem~\ref{T:Recovery_incomplete} provides sufficient conditions for the relaxed problem in \eqref{E:SparseAdj_l00_onlysomeeig} to recover the sparsest graph, even when incomplete information about the eigenvectors is available. In practice it is observed that for smaller number $K$ of known spectral templates the value of $\eta_{\bbP}$ in \eqref{E:condition_recovery_incomplete} tends to be larger, indicating a less favorable setting for recovery. This observation is aligned with the results obtained in practice; see Fig.~\ref{F:num_exp_2}(c).

To state results similar to those in Theorem~\ref{T:Recovery_incomplete} but for the recovery of normalized Laplacians, we define $\tilde{\bbU}_K := \tilde{\bbV}_K \odot  \tilde{\bbV}_K$ where $ \tilde{\bbV}_K := [\bbv_2, \bbv_3, \ldots, \bbv_K]$ and define the matrices
\begin{equation}\label{E:def_T_matrices}
	\bbT_1 := 
	\begin{bmatrix}
		\bbI - \tilde{\bbU}_K \tilde{\bbU}_K^\dag \\
		\bbI_\ccalD \\
		\bbB \\
		\bbzero_{NK \times N^2} \\
	\end{bmatrix}^T, \quad
	\bbT_2 := 
	\begin{bmatrix}
		\tilde{\bbU}_K \tilde{\bbU}_K^\dag - \bbI \\
		\bbzero_{N \times N^2} \\
		\bbzero_{{N \choose 2} \times N^2} \\
		\bbI \otimes V_K^T \\
	\end{bmatrix}^T,
\end{equation}
and $\bbT := [\bbT_1^T , \bbT_2^T]^T$. Under the assumption that the first eigenvector (i.e., the one whose associated eigenvalue is zero) is among the $K$ eigenvectors known, the following result holds. The proof -- here omitted -- follows the same steps as those in Theorem~\ref{T:Recovery_incomplete}.

\begin{mytheorem}\label{T:Recovery_incomplete_Laplacian}
	Whenever $\ccalS = \ccalS_{\rm L}$ and assuming problem \eqref{E:SparseAdj_l00_onlysomeeig} is feasible, $\bar{\bbS}^*_1 = \bbS^*_0$ if the two following conditions are satisfied:\\
	L-1) $\rank([{\bbT_1}_\ccalJ^T, \bbT_2^T]) = |\ccalJ| + N^2$ ; and \\
	L-2) There exists a constant $\delta > 0$ such that 
	\begin{equation}\label{E:condition_recovery_incomplete_laplacian}
		\eta_{\bbT} := \| \bbUpsilon_{\ccalJ^c}(\delta^{-2} \bbT \bbT^T + \bbUpsilon_{\ccalJ^c}^T \bbUpsilon_{\ccalJ^c})^{-1}\bbUpsilon_{\ccalJ}^T \|_{M(\infty)}<1.
	\end{equation}
\end{mytheorem}


Notice that scenarios that combine the settings in Sections~\ref{Ss:ShiftRecImperfEig} and~\ref{Ss:ShiftRecIncompEig}, i.e. where the knowledge of the $K$ templates is imperfect, can be handled by combining the formulations in \eqref{E:SparseAdj_l1_obj_noisy_matrix_v2} and \eqref{E:SparseAdj_l00_onlysomeeig}. This can be achieved upon implementing the following modifications to \eqref{E:SparseAdj_l00_onlysomeeig}: considering the shift $\bbS'$ as a new optimization variable, replacing the first constraint in \eqref{E:SparseAdj_l00_onlysomeeig} with $\bbS'=\bbS_{\bar{K}}+{\textstyle \sum_{k=1}^{K}} \lambda_k \bbv_k \bbv_k^T$, and adding $d(\bbS,\bbS') \leq \epsilon$ as a new constraint [cf.~\eqref{E:SparseAdj_l1_obj_noisy_matrix_v2}].


\section{Numerical experiments}\label{S:Simulations}



We test the proposed topology inference methods on different synthetic and real-world graphs. A comprehensive performance evaluation is carried out whereby we: (i)~investigate the recovery of both adjacency and normalized Laplacian matrices; (ii)~corroborate our main theoretical findings; (iii)~assess the impact of imperfect information in the recovery;  (iv)~carry out comparisons with state-of-the-art methods; and (v)~illustrate how our framework can be used to promote sparsity on a given network.

\subsection{Topology inference from noiseless templates}\label{Ss:Sim_Top_inf_noiseless_templates}

Consider Erd\H{o}s-R\'enyi (ER) graphs \cite{bollobas1998random} of varying size $N \in \{10, 20, \dots, 50\}$ and different edge-formation probabilities $p \in \{0.1, 0.2, \ldots, 0.9\}$. For each combination of $N$ and $p$ we generate 100 graphs and try to recover their adjacency $\bbA$ and normalized Laplacian $\bbL$ matrices from the corresponding spectral templates $\bbV$. In Fig.~\ref{F:num_exp}(a) we plot the proportion of instances where the corresponding optimization problems -- problem \eqref{E:general_problem} for $\ccalS = \ccalS_\mathrm{A}$ and $\ccalS = \ccalS_\mathrm{L}$ -- have singleton feasibility sets. Notice that multiple solutions are more frequent when the expected number of neighbors of a given node is close to either $1$ or $N$. For intermediate values of $p$, the rank of both ${\bbW}_{\ccalD}$ and $\bbU$ is typically $N-1$, guaranteeing a single feasible point (cf. Proposition~\ref{P:Feasibility_singleton}). Using the same set of graphs that those in Fig.~\ref{F:num_exp}(a), Fig.~\ref{F:num_exp}(b) shows the recovery rate when solving the iteratively re-weighted problem \eqref{E:SparseAdj_wl11_obj} for both the adjacency (top) and the normalized Laplacian (bottom). As expected, the rates in Fig.~\ref{F:num_exp}(b) dominate those in Fig.~\ref{F:num_exp}(a) since every instance with a unique feasible point is recovered successfully. Moreover, the improved rates observed in Fig.~\ref{F:num_exp}(b) are indicative of the beneficial effect that the weighted $\ell_1$ norm objective has in the recovery.

\begin{figure*}
	\centering
	\begin{minipage}[c]{.33\textwidth}
		\includegraphics[width=\textwidth]{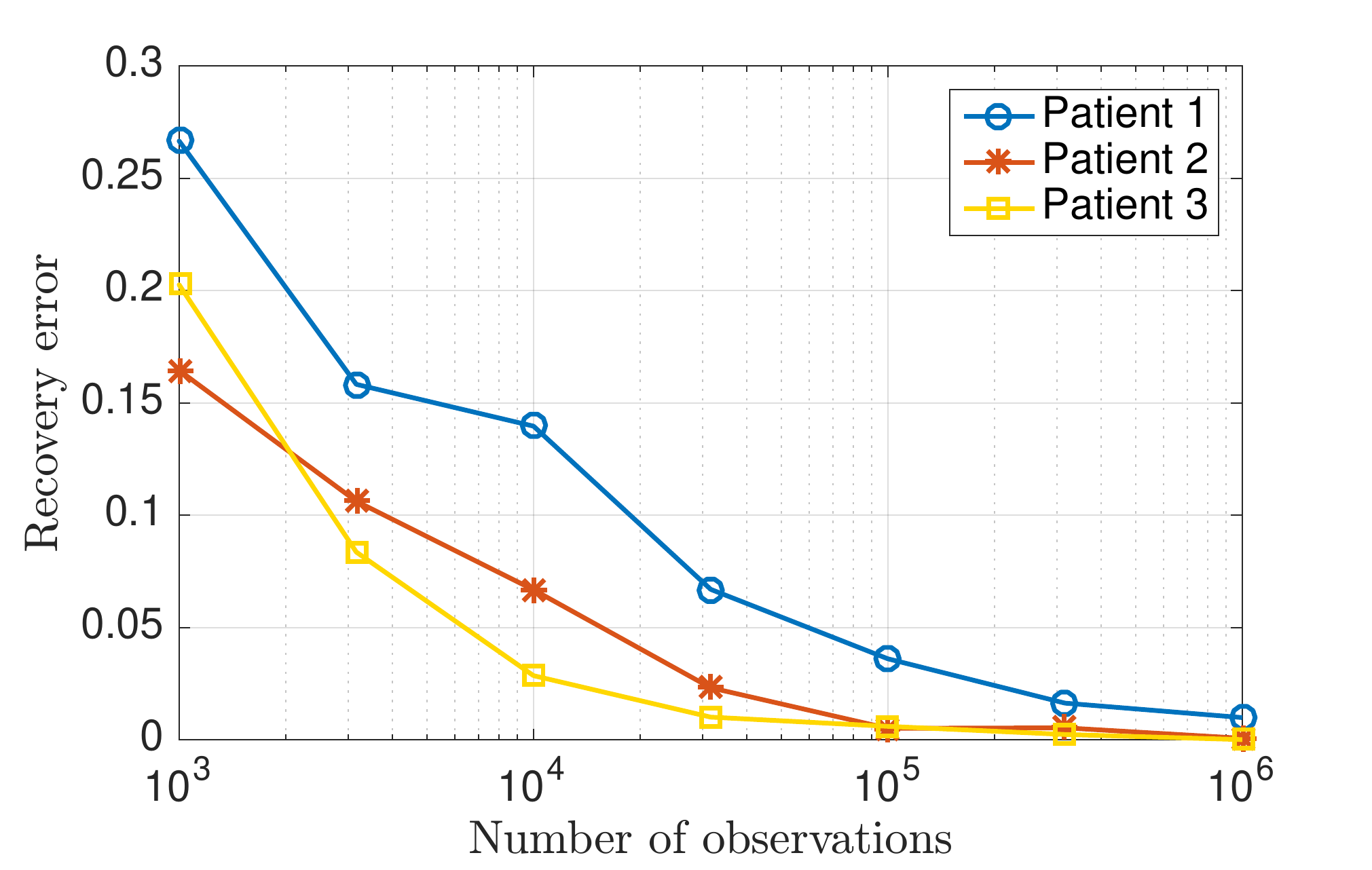}
		\centering{\small (a)}
	\end{minipage}%
	\begin{minipage}[c]{.33\textwidth}
		\includegraphics[width=\textwidth]{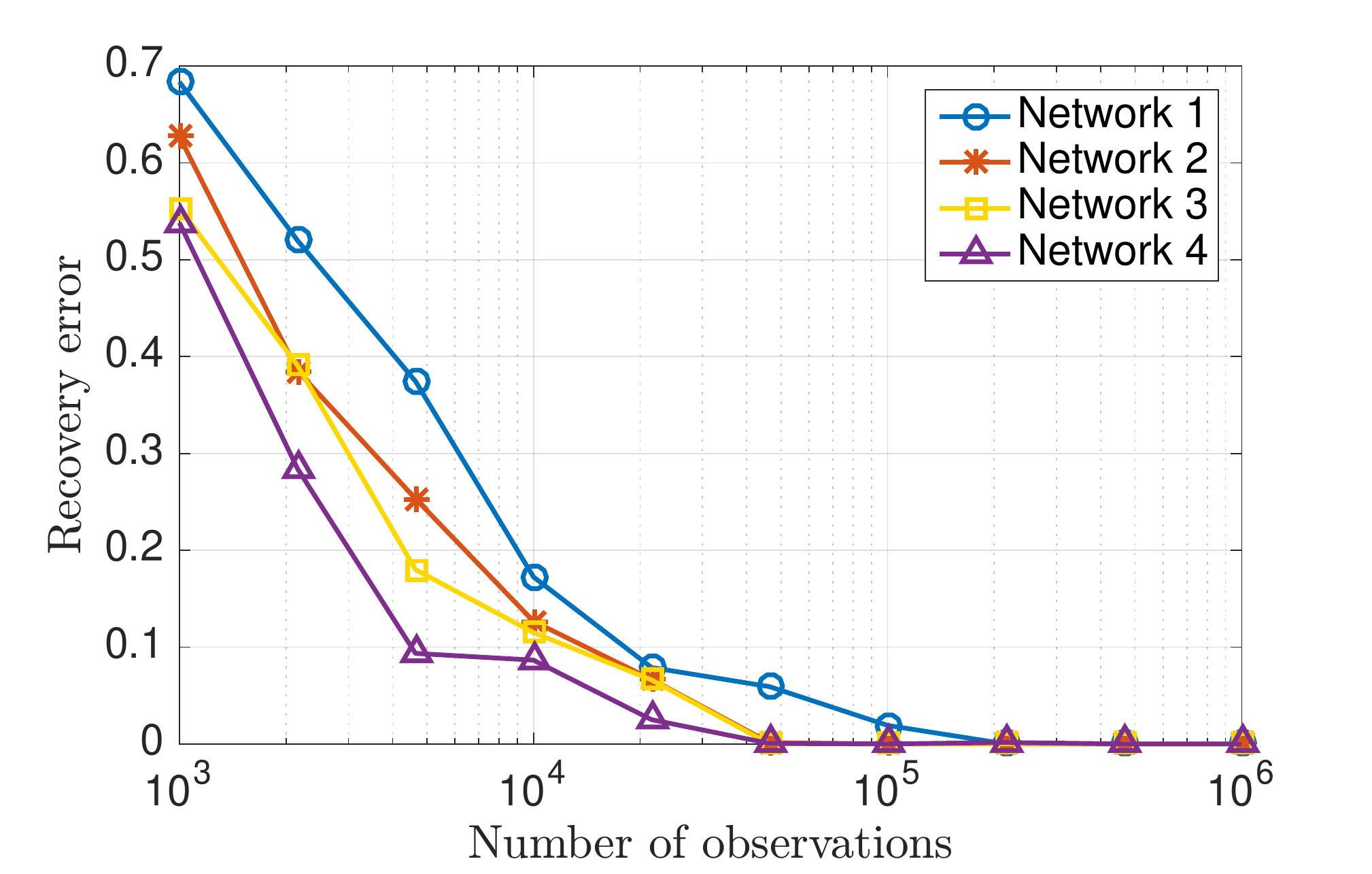}
		\centering{\small (b)}
	\end{minipage}%
	\begin{minipage}{.33\textwidth}
		\includegraphics[width=\textwidth]{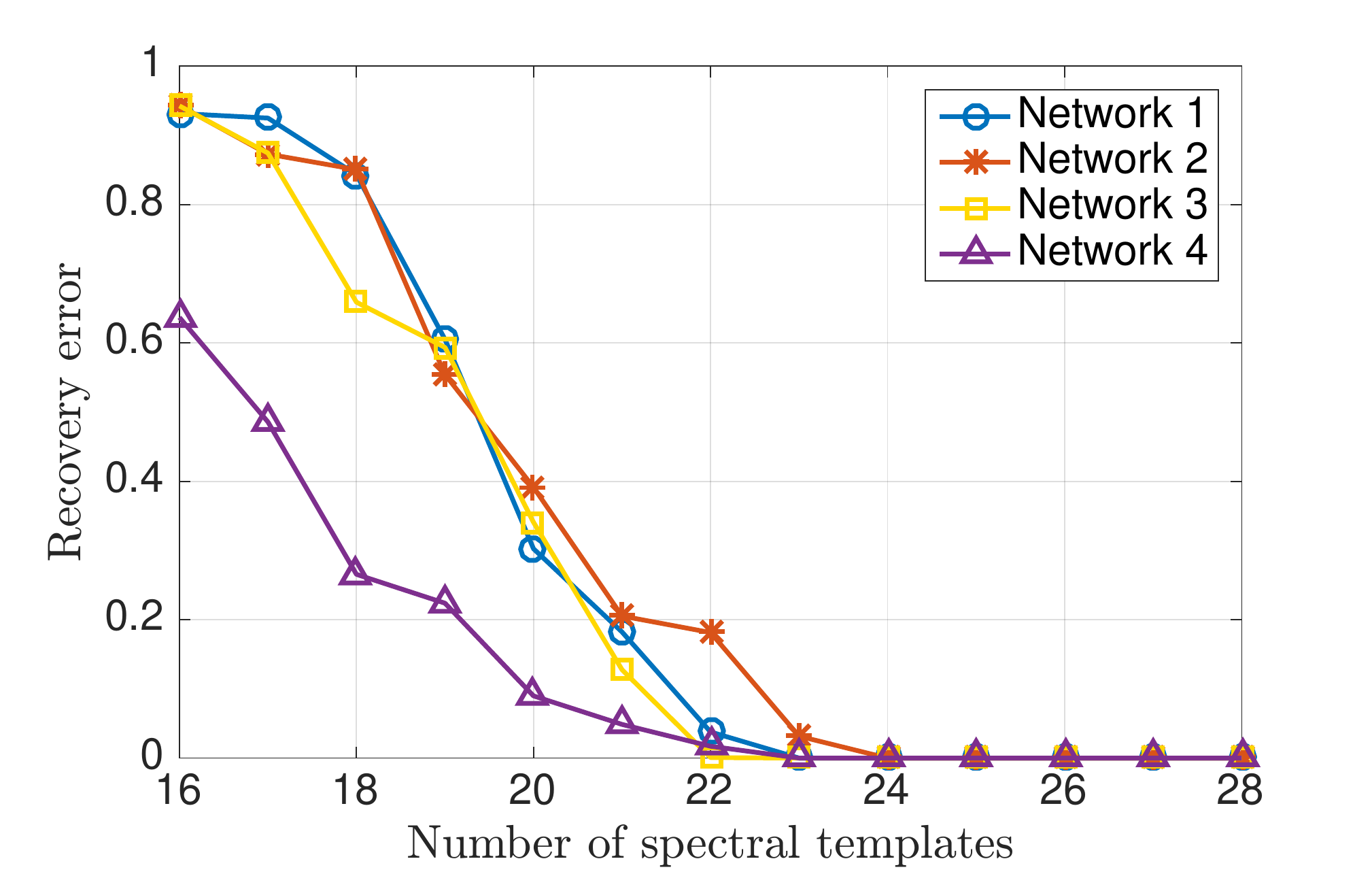}
		\centering{\small (c)}
	\end{minipage}
	\caption{ (a) Brain graph recovery error for three patients as a function of the number of signals observed in the estimation of the spectral templates.  (b)~Recovery error for four social networks as a function of the number of signals observed in the estimation of the spectral templates. (c) Recovery error for four social networks (with $N=32$ nodes) as a function of $K$, the number of spectral templates that are known.}
	\vspace{-0.15in}
	\label{F:num_exp_2}
\end{figure*}

As indicated by Proposition~\ref{P:Feasibility_singleton}, the rate of recovery is intimately related to the ranks of ${\bbW}_{\ccalD}$ and $\bbU$ for the adjacency and normalized Laplacian cases, respectively. Fig.~\ref{F:num_exp}(c) further illustrates this relation via a histogram of the rank of $\bbU$ for the 100 graphs with $N=10$ and $p=0.2$. For more than half of the instances, the rank of $\bbU$ is equal to 9 (blue bar) and, as stated in Proposition~\ref{P:Feasibility_singleton}, for all these graphs there is a unique feasible point (yellow bar) that is successfully recovered (cyan bar). We see that, as the rank of $\bbU$ degrades, uniqueness is no longer guaranteed but for most cases the true graph can still be recovered following the iteratively re-weighted scheme proposed. Only in 8 of the cases where $\mathrm{rank}(\bbU)<9$ the recovery was not successful, entailing a recovery rate of 0.92, as reported in the corresponding entry ($N=10$, $p=0.2$) of the bottom plot in Fig.~\ref{F:num_exp}(b).

Finally, in order to corroborate the conditions for noiseless recovery stated in Theorem~\ref{T:Recovery}, we draw ER random graphs of size $N \!=\! 20$ and edge-formation probability $p \! = \! 0.25$. For each graph, we make sure that the associated $\bbW_{\ccalD}$ matrix has rank strictly smaller than $N-1$ (to rule out the cases where the feasible set is a singleton), and that condition \emph{A-1)} in Theorem~\ref{T:Recovery} is satisfied. In Fig.~\ref{F:num_exp}(d) we plot the number of successes and failures in recovering the adjacency as a function of $\psi_\bbR$ in \eqref{E:condition_recovery_noiseless}. We consider 1000 realizations and for each of them the constant $\delta$ in \eqref{E:condition_recovery_noiseless} is chosen to minimize $\psi_\bbR$. Fig.~\ref{F:num_exp}(d) clearly depicts the result of Theorem~\ref{T:Recovery} in that, for all cases where $\psi_\bbR<1$, relaxation \eqref{E:SparseAdj_l01_obj} achieves perfect recovery. Equally important, it is clear that the bound stated in \eqref{E:condition_recovery_noiseless} is tight since a large proportion of the realizations with $\psi_\bbR$ equal to 1 or just above this value lead to failed recoveries.

\subsection{Topology inference from noisy and incomplete templates}\label{Ss:Sim_Top_inf_noisy_templates}

We consider the identification of unweighted and undirected graphs corresponding to human brains~\cite{hagmann2008mapping}. Each graph consists of $N=66$ nodes, which represent brain regions of interest (ROIs). An edge between two ROIs exists if the density of anatomical connections is greater than a threshold, which is chosen as the largest one that entails a connected graph~\cite{hagmann2008mapping}. We test the recovery from noisy spectral templates $\hat{\bbV}$ [cf.~\eqref{E:SparseAdj_l1_obj_noisy_matrix_v2}] obtained from sample covariances of synthetic signals generated through diffusion processes (cf.~Section~\ref{S:prelim_problem}). 
These processes are modeled by graph filters of random degree between $3$ and $7$, and with independent and normally distributed coefficients.
Denoting by $\hat{\bbV}_i$ the noisy spectral templates of patient $i \! \in \! \{1, 2, 3\}$ and by $\hat{\bbA}_i$ the adjacency matrices recovered, Fig.~\ref{F:num_exp_2}(a) plots the recovery error as a function of the number of signals observed in the computation of the sample covariance. The error is quantified as the proportion of edges misidentified, i.e., $\|\bbA_i \!-\! \hat{\bbA}_i\|_0 / \| \bbA_i \|_0$, and each point in Fig.~\ref{F:num_exp_2}(a) is the average across 50 realizations. 
Notice that for an increasing number of observed signals we see a monotonous decrease in the recovery error. For example, when going from $10^4$ to $10^5$ observations the error is (approximately) divided by seven, when averaged across patients. This is reasonable since a larger number of observations gives rise to a more reliable estimate of the covariance matrix entailing less noisy spectral templates. 
Traditional methods like graphical lasso \cite{GLasso2008} fail to recover $\bbS$ from the sample covariance of filtered white signals. 
For example, when signals are generated using a filter of the form $\bbH = h_0 \bbI + h_1 \bbS$, graphical lasso performs significantly worse than the method based on spectral templates. More precisely, when $10^5$ signals are observed, the recovery error of graphical lasso averaged over 50 realizations and with optimal tuning parameters is 0.303, 0.350, and 0.270 for patients 1, 2, and 3, respectively. Such errors are between 5 and 50 times larger than those reported in Fig.~\ref{F:num_exp_2}(a). Further comparisons of our method with graphical lasso and other existing alternatives are provided in Section~\ref{Ss:Sim_performance_comparison}. 

\begin{figure}[t]
	\centering
	\includegraphics[width=0.44\textwidth]{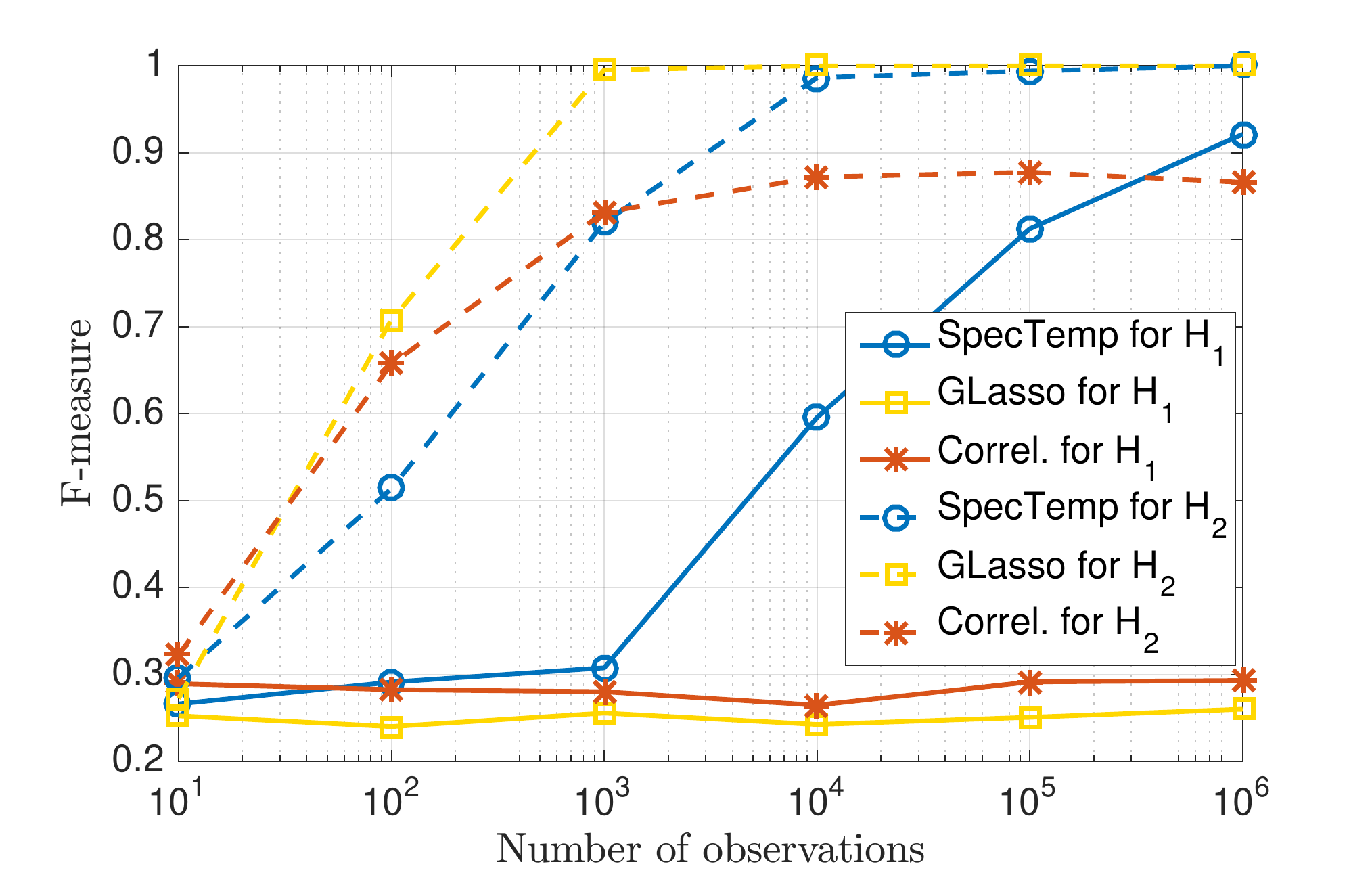}
	\vspace{-0.1in}	
	\caption{Performance comparison between spectral templates (SpecTemp), graphical lasso, and correlation-based recovery. For general filters, SpecTemp outperforms the other two.}
	\vspace{-0.15in}
	\label{F:comparison_traditional}
\end{figure}

We repeat the previous experiment on four social networks defined on a common set of $N=32$ nodes, which represent students from the University of Ljubljana\footnote{Access to the data and additional details are available at \url{http://vladowiki.fmf.uni-lj.si/doku.php?id=pajek:data:pajek:students}}.
Links for each of the networks capture different types of interactions among the students, and were built by asking each student to select a group of preferred college mates for different situations, e.g., to discuss a personal issue or to invite to a birthday party (see footnote 2 for further details). The considered graphs are unweighted and symmetric, and the edge between $i$ and $j$ exists if either student $i$ picked $j$ in the questionnaire or vice versa. As done for the brain graphs, we test the recovery performance for noisy spectral templates $\hat{\bbV}$ obtained from sample covariances. Fig.~\ref{F:num_exp_2}(b) plots the reconstruction error as a function of the number of observed signals for the different networks studied. As was observed in Fig.~\ref{F:num_exp_2}(a), we see a monotonous decrease in recovery error for all the analyzed networks.


\begin{table*}
	\centering
	\caption{Performance comparison between spectral templates (SpecTemp), Kalofolias \cite{Kalofolias2016inference_smoothAISTATS16}, and Dong etal \cite{DongLaplacianLearning}.}
	\vspace{-0.1in}
	\begin{tabular}{r c c c c c c c c c c c c} 
		& & \multicolumn{3}{c}{Inverse Laplacian} & & \multicolumn{3}{c}{Diffusion} & & \multicolumn{3}{c}{Exponential} \\[1pt]
		\cline{3-5}  \cline{7-9}  \cline{11-13} \\[-6pt]
		& & SpecTemp & Kalofolias & Dong etal & & SpecTemp & Kalofolias & Dong etal & & SpecTemp & Kalofolias & Dong etal \\[1pt]
		\hline
		{\bf Erd\H{o}s-R\'enyi} & & & & & & & & & & & & \\
		F-measure &  & {\bf 0.896}  &  0.791  &  0.818 &  & {\bf 0.924}  &  0.868  &  0.828 &  & {\bf 0.703}  &  0.651  &  0.667 \\
		edge error &  & {\bf 0.108}  &  0.152  &  0.168 &  & {\bf 0.071}  &  0.149  &  0.177 &  & {\bf 0.276}  &  0.318  &  0.332 \\
		degree error & & {\bf 0.058}  &  0.071  &  0.105 & & {\bf 0.040}  &  0.055  &  0.111 & & {\bf 0.162}   & 0.201  &  0.222 \\
		{\bf Barab\'asi-Albert} & & & & & & & & & & & & \\
		F-measure & & {\bf 0.926}  &  0.855  &  0.873 & & {\bf 0.945}  &  0.845  &  0.894 & & {\bf 0.814}  &  0.732  &  0.798 \\
		edge error & & {\bf 0.143}  &  0.173  &  0.209 & & {\bf 0.135}  &  0.154  &  0.235 & & {\bf 0.310}  &  0.314  &  0.393 \\
		degree error & & {\bf 0.108}  &  0.124  &  0.169 & & 0.109  &  {\bf 0.092}  &  0.188 & & {\bf 0.240}  &  0.244  &  0.282 \\
		\hline
	\end{tabular}
	\vspace{-0.1in}
	\label{Tab:comparison_methods}
\end{table*}

Finally, we illustrate the recovery performance in the presence of incomplete spectral templates by solving \eqref{E:SparseAdj_l00_onlysomeeig} for the four networks in Fig.~\ref{F:num_exp_2}(b). More specifically, in Fig.~\ref{F:num_exp_2}(c) we plot the recovery error as a function of the number $K$ of eigenvectors available. Each point in the plot is the average across 50 realizations in which different $K$ eigenvectors were selected from the $N\!=\!32$ possible ones. As expected, the performance for all four networks improves with the number of spectral templates known. The performance improvement is sharp and precipitous going from a large error of over $0.85$ for three of the networks when 17 spectral templates are known to a perfect recovery for all the networks when 24 eigenvectors are given. Moreover, notice that network $4$ is consistently the easiest to identify both for noisy [cf. Fig.~\ref{F:num_exp_2}(b)] and incomplete [cf. Fig.~\ref{F:num_exp_2}(c)] spectral templates. For example, when given 19 spectral templates the error associated with network $4$ is $0.224$ whereas the average across the other three networks is $0.584$. This hints towards the fact that some graphs are inherently more robust for identification when given imperfect spectral templates. A formal analysis of this phenomenon is left as future work.

\subsection{Performance comparison}\label{Ss:Sim_performance_comparison}

We compare the performance of the presented method based on spectral templates (we refer to it as SpecTemp for conciseness) with established statistical approaches as well as recent GSP-based algorithms.

\noindent\textbf{Comparison with established methods.} We analyze the performance of SpecTemp in comparison with two widely used methods, namely, (thresholded) correlation~\cite[Ch. 7.3.1]{kolaczyk2009book}  and graphical lasso \cite{GLasso2008}. 
Our goal is to recover the adjacency matrix of an undirected and unweighted graph with no self-loops from the observation of filtered graph signals. 
For the implementation of SpecTemp, we use the eigendecomposition of the sample covariance of the observed signals in order to extract noisy spectral templates $\hat{\bbV}$. We then solve problem \eqref{E:SparseAdj_l1_obj_noisy_matrix_v2} for $\ccalS = \ccalS_{\mathrm{A}}$, where $\epsilon$ is selected as the smallest value that admits a feasible solution. We include as a priori knowledge that each node has at least one neighbor. For the correlation-based method, we keep the absolute value of the sample correlation of the observed signals, force zeros on the diagonal and set all values below a certain threshold to zero. This threshold is determined during a training phase, as explained in more detail in the next paragraph. Lastly, for graphical lasso we follow the implementation in \cite{GLasso2008} based on the sample covariance and select the tuning parameter $\rho$ (see \cite{GLasso2008}) during the training phase. We then force zeros on the diagonal and keep the absolute values of each entry. Leveraging that the sought graphs are unweighted, for SpecTemp and graphical lasso a fixed threshold of 0.3 is used so that, after recovery, every edge with weight smaller than the threshold is set to zero.

\begin{figure}[t]
	\centering
	\includegraphics[width=0.41\textwidth]{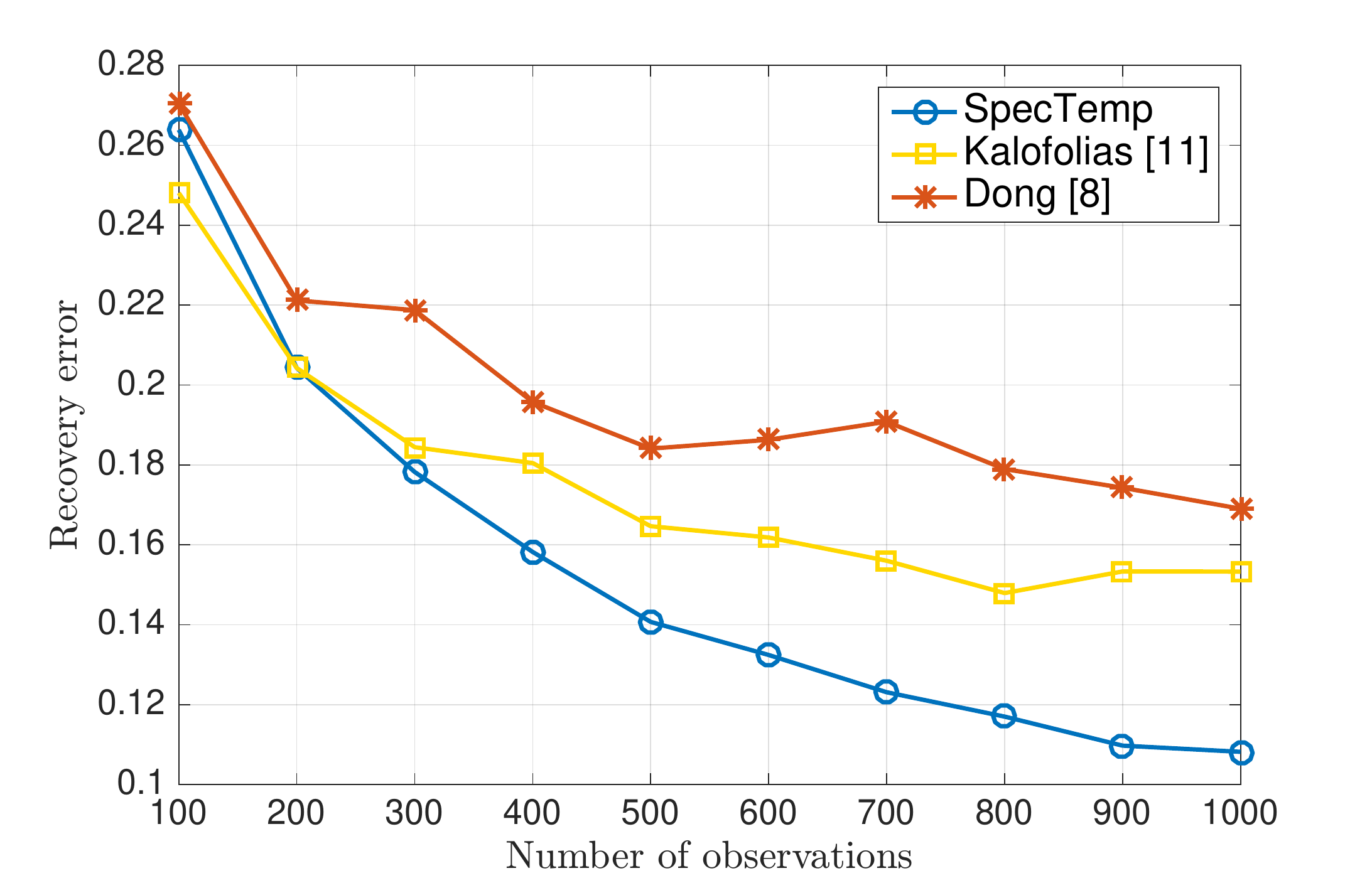}
	\vspace{-0.1in}	
	\caption{Comparison of edge recovery error as a function of the number of signals observed for SpecTemp, Kalofolias \cite{Kalofolias2016inference_smoothAISTATS16}, and Dong etal\cite{DongLaplacianLearning}.}
	\vspace{-0.15in}
	\label{F:comparison_kalo_dong}
\end{figure}

We test the recovery of adjacency matrices $\bbS \! = \! \bbA$ of ER graphs with $N \! = \! 20$ nodes and edge probability $p \! = \! 0.2$. We vary the number of observed signals from $10^1$ to $10^6$ in powers of $10$. Each of these signals is generated by passing white Gaussian noise through a graph filter $\bbH$. Two different types of filters are considered. 
As a first type we consider a \emph{general} filter $\bbH_1 = \bbV \diag(\widehat{\bbh}_1) \bbV^T$, where the entries of $\widehat{\bbh}_1$ are independent and chosen randomly between $0.5$ and $1.5$. 
The second type is a \emph{specific} filter of the form $\bbH_2 = (\delta \bbI + \bbS)^{-1/2}$, where the constant $\delta$ is chosen so that $\delta \bbI + \bbS$ is positive definite to ensure that $\bbH_2$ is real and well-defined. Following the discussion in Section~\ref{S:prelim_problem}, this implies that the precision matrix of the filtered signals is given by $\bbC_x^{-1} = \bbH_2^{-2} = \delta \bbI + \bbS$, which coincides with the desired GSO $\bbS$ in the off-diagonal elements. 
For each combination of filter type and number of observed signals, we generate 10 ER graphs that are used for training and 20 ER graphs that are used for testing. Based on the 10 training graphs, the optimal threshold for the correlation method and parameter $\rho$ for graphical lasso are determined and then used for the recovery of the 20 testing graphs. Given that for SpecTemp we are fixing $\epsilon$ beforehand, no training is required.

\begin{figure*}[t]
	\centering
	\input{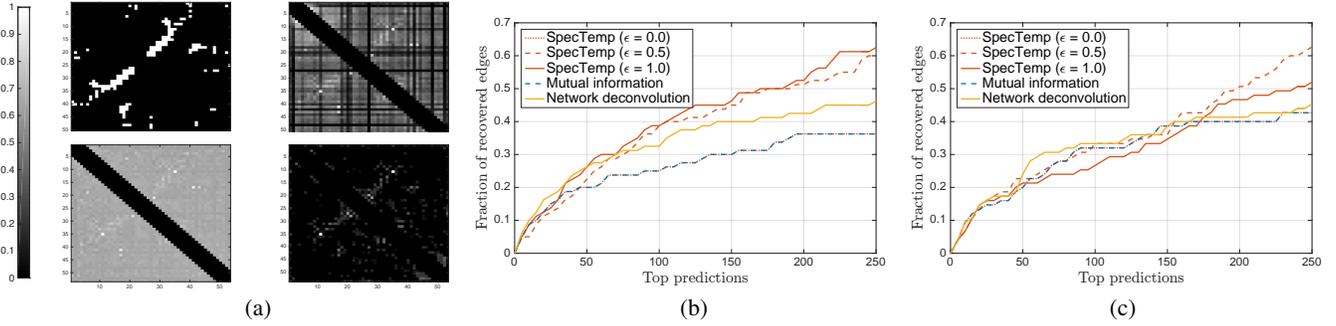}	
	\vspace{-0.25in}
	\caption{(a) Real and inferred contact networks between amino-acid residues for protein BPT1 BOVIN. Ground truth contact network (top left), mutual information of the co-variation of amino-acid residues (top right), contact network inferred by network deconvolution (bottom left), contact network inferred by our method based on spectral templates (bottom right). (b) Fraction of the real contact edges between amino-acids recovered for each method as a function of the number of edges considered. (c) Counterpart of (b) for protein YES HUMAN.}
	\vspace{-0.1in}
	\label{F:network_sparsification}
\end{figure*}

As figure of merit we use the F-measure \cite{manning2008introduction}, i.e. the harmonic mean of edge precision and edge recall, that solely takes into account the support of the recovered graph while ignoring the weights. In Fig.~\ref{F:comparison_traditional} we plot the performance of the three methods as a function of the number of filtered graph signals observed for filters $\bbH_1$ and $\bbH_2$, where each point is the mean F-measure over the 20 testing graphs. 

When considering a general graph filter $\bbH_1$ SpecTemp clearly outperforms the other two. For instance, when $10^5$ signals are observed, our average F-measure is $0.81$ while the measures for correlation and graphical lasso are $0.29$ and $0.25$, respectively. Moreover, of the three methods, our approach is the only consistent one, i.e., achieving perfect recovery with increasing number of observed signals. Although striking at a first glance, the deficient performance of graphical lasso was expected. For general filters $\bbH_1$, the precision matrix is given by $\bbC_x^{-1} = \bbH_1^{-2}$ which in general is neither sparse nor shares the support of $\bbS$, the GSO to be recovered. When analyzing the specific case of graph filters $\bbH_2$, where the precision matrix exactly coincides with the desired graph-shift operator, graphical lasso outperforms both our method and the correlation-based method. This is not surprising since graphical lasso was designed for the recovery of sparse precision matrices. Notice however that for large number of observations SpecTemp, without assuming any specific filter model, also achieves perfect recovery and yields an F-measure equal to 1.

\noindent\textbf{Comparison with GSP methods.} We compare the recovery using SpecTemp with the algorithms in \cite{DongLaplacianLearning} and \cite{Kalofolias2016inference_smoothAISTATS16}, both methods designed to identify the (combinatorial) Laplacian of a graph when given a set of smooth graph signals. Small modifications can be made to our framework to accommodate this setting, thus permitting a fair comparison. More precisely, in solving \eqref{E:SparseAdj_l1_obj_noisy_matrix_v2} we use the set of admissible shifts given by 
\begin{align}\label{E:def_S_combinatorial_laplacian}
	\ccalS_{\rm L_c}  :=  \{ \bbS \, | \, S_{ij} \leq 0   \,\, \text{for} \,\,& i\!\neq\! j,  \;\;\bbS\! \in \! \ccalM_{+}^N, \;\; \bbS \bbone = \bbzero \}.
\end{align}
Moreover, in order to account for the smoothness of the observed signals in the unknown graph we sort the eigenvectors $\hbv_k$ of the sample covariance in increasing order of their corresponding eigenvalues, and we require the recovered eigenvalues $\bblambda$ to satisfy $\lambda_{i} \geq \lambda_{i+k} + \delta$ for all $i$, and fixed $k$ and $\delta$. In this way, we assign the frequencies with larger presence in the observed signals to low eigenvalues in the recovered Laplacian. Unless otherwise noted, we set $\delta = 0.1$ and $k = 3$.

We compare the three methods of interest on two different types of graphs and three different signal generation models. We consider the recovery of the Laplacian $\bbS = \bbL$ of ER graphs with $N \! = \! 20$ nodes and edge probability $p \! = \! 0.3$ as well as Barab\'asi-Albert preferential attachment graphs \cite{bollobas1998random} with $N \! = \! 20$ generated from $m_0 = 4$ initially placed nodes, where each new node is connected to $m=3$ existing ones. Following \cite{Kalofolias2016inference_smoothAISTATS16} we consider three models for smooth graph signals: i) multivariate normal signals with covariance given by the pseudo-inverse of $\bbL$, i.e., $\bbx_1 \sim \ccalN(\bbzero, \bbL^{\dag})$; ii) white signals filtered through an autoregressive (diffusion) process, that is $\bbx_2 = (\bbI + \bbL)^{-1} \bbw$, where $\bbw\sim \ccalN(\bbzero, \bbI)$; and iii) white signals passed through an exponential filter, $\bbx_3 = \exp(-\bbL) \bbw$. For each of the six settings considered (two graphs combined with three signal types) we generate 10 training graphs, 100 testing graphs, and for every graph we generate 1000 graph signals. The training set is used to set the parameters in \cite{DongLaplacianLearning} and \cite{Kalofolias2016inference_smoothAISTATS16}, and in our case it serves the purpose of selecting the best $\epsilon$ [cf.~\eqref{E:SparseAdj_l1_obj_noisy_matrix_v2}]. To increase the difficulty of the recovery task, every signal $\bbx$ is perturbed as $\hbx = \bbx + \sigma \, \bbx \circ \bbz$, for $\sigma = 0.1$ and where each entry of $\bbz$ is an independent standard normal random variable. We focus on three performance measures, namely, the F-measure as explained in the previous experiment, the $\ell_2$ relative error of recovery of the edges, and the $\ell_2$ relative error of recovery of the degrees. 
The performance achieved by each method in the testing sets is summarized in Table~\ref{Tab:comparison_methods}. In all but one case, our method attains the largest F-measures and the smallest errors for all the graphs and signal types considered. 

Finally, for the particular cases of ER graphs and signals $\bbx_1$ (inverse Laplacian), we replicate the above procedure varying the number of observed signals $P$ from $100$ to $1000$. For SpecTemp, we increase $k$ when the number of observations decreases to account for the noisier ordering of the eigenvectors in the sample covariance. In this experiment we use $k=5$ for $P \leq 400$, $k=4$ for $400<P<800$ and $k=3$ for $P \geq 800$. In Fig.~\ref{F:comparison_kalo_dong} we plot the associated $\ell_2$ edge recovery errors. Notice that for small number of observations, the method in \cite{Kalofolias2016inference_smoothAISTATS16} outperforms SpecTemp whereas the opposite is true when more signals are observed. This can be attributed to the fact that SpecTemp assumes no specific model on the smoothness of the signal, thus, when enough signals are observed our more agnostic, data-driven approach exhibits a clear performance advantage.

\subsection{Network sparsification}\label{Ss:Sim_network_sparsification}


With reference to the network sparsification problem outlined in Section \ref{sssec_sparsify}, our goal here is to identify the structural properties of proteins from a mutual information graph of the co-variation between the constitutional amino-acids \cite{Marks2011proteins}; see \cite{FeiziNetworkDeconvolution} for details. For example, for a particular protein, we want to recover the structural graph in the top left of Fig.~\ref{F:network_sparsification}(a) when given the graph of mutual information in the top right corner. Notice that the structural contacts along the first four sub-diagonals of the graphs were intentionally removed to assess the capability of the methods in detecting the contacts between distant amino-acids. The graph recovered by network deconvolution \cite{FeiziNetworkDeconvolution} is illustrated in the bottom left corner of Fig.~\ref{F:network_sparsification}(a) whereas the one recovered using SpecTemp is depicted in the bottom right corner of the figure. Comparing both recovered graphs, SpecTemp leads to a sparser graph that follows more closely the desired structure to be recovered. To quantify this latter assertion, in Fig.~\ref{F:network_sparsification}(b) we plot the fraction of the real contact edges recovered for each method as a function of the number of edges considered, as done in \cite{FeiziNetworkDeconvolution}. 
For example, if for a given method the 100 edges with largest weight in the recovered graph contain $40\%$ of the edges in the ground truth graph we say that the 100 top edge predictions achieve a fraction of recovered edges equal to $0.4$.
As claimed in \cite{FeiziNetworkDeconvolution}, network deconvolution improves the estimation when compared to raw mutual information data. Nevertheless, from Fig.~\ref{F:network_sparsification}(b) it follows that SpecTemp outperforms network deconvolution. Notice that when $\epsilon = 0$ [cf.~\eqref{E:SparseAdj_l1_obj_noisy_matrix_v2}] we are forcing the eigenvectors of $\bbS$ to coincide exactly with those of the matrix of mutual information $\bbS'$. However, since $\bbS'$ is already a valid adjacency matrix, we end up recovering $\bbS = \bbS'$. By contrast, for larger values of $\epsilon$ the additional flexibility in the choice of the eigenvectors allows us to recover shifts $\bbS$ that more closely resemble the ground truth. For example, when considering the top $200$ edges, the mutual information and the network deconvolution methods recover $36\%$ and $43\%$ of the desired edges, respectively, while our method for $\epsilon \! = \! 1$ achieves a recovery of $53\%$. In Fig.~\ref{F:network_sparsification}(c) we present this same analysis for a different protein and similar results can be appreciated.


\section{Conclusions}\label{S:Conclusions}


With $\bbS=\bbV\bbLambda\bbV^T$ being the shift operator associated with the graph $\ccalG$, we studied the problem of identifying $\bbS$ (hence the topology of $\ccalG$) using a two-step approach under which we first obtain $\bbV$, and then use $\bbV$ as input to find $\bbLambda$. The problem of finding $\bbLambda$ given $\bbV$ was formulated as a sparse recovery optimization. Efficient algorithms based on convex relaxations were developed, and theoretical conditions under which exact and robust recovery is guaranteed were derived for the cases where $\bbS$ represents the adjacency or the normalized Laplacian of $\ccalG$. 
%
%
In identifying $\bbV$, our main focus was on using as input a set of graph signal realizations. Under the assumption that such signals resulted from diffusion dynamics on the graph or, equivalently, that they were stationary in $\bbS$, it was shown that $\bbV$ could be estimated from the eigenvectors of the sample covariance of the available set. As a consequence, several well-established methods for topology identification based on local and partial correlations can be viewed as particular instances of the approach here presented.
The practical relevance of the proposed schemes and the gains relative to existing alternatives were highlighted carrying out numerical tests with synthetic and real-world graphs.


\bibliographystyle{IEEEtran}
%
\bibliography{citations}

\begin{thebibliography}{10}
\providecommand{\url}[1]{#1}
\csname url@samestyle\endcsname
\providecommand{\newblock}{\relax}
\providecommand{\bibinfo}[2]{#2}
\providecommand{\BIBentrySTDinterwordspacing}{\spaceskip=0pt\relax}
\providecommand{\BIBentryALTinterwordstretchfactor}{4}
\providecommand{\BIBentryALTinterwordspacing}{\spaceskip=\fontdimen2\font plus
\BIBentryALTinterwordstretchfactor\fontdimen3\font minus
  \fontdimen4\font\relax}
\providecommand{\BIBforeignlanguage}[2]{{%
\expandafter\ifx\csname l@#1\endcsname\relax
\typeout{** WARNING: IEEEtran.bst: No hyphenation pattern has been}%
\typeout{** loaded for the language `#1'. Using the pattern for}%
\typeout{** the default language instead.}%
\else
\language=\csname l@#1\endcsname
\fi
#2}}
\providecommand{\BIBdecl}{\relax}
\BIBdecl

\bibitem{SSAMGMAR_ssp16}
S.~Segarra, A.~G. Marques, G.~Mateos, and A.~Ribeiro, ``Network topology
  identification from spectral templates,'' in \emph{IEEE Wrkshp. Statistical
  Signal Process. (SSP)}, Palma de Mallorca, Spain, Jun. 26-29, 2016.

\bibitem{SSAMGMAR_Asilomar16}
------, ``Network topology identification from imperfect spectral templates,''
  in \emph{Asilomar Conf. on Signals, Systems, and Computers}, Pacific Grove,
  CA, Nov. 6-9, 2016.

\bibitem{barrat2012book}
A.~Barrat, M.~Barth\'{e}lemy, and A.~Vespignani, \emph{Dynamic Processes on
  Complex Networks}.\hskip 1em plus 0.5em minus 0.4em\relax Cambridge, UK:
  Cambridge University Press, 2012.

\bibitem{kolaczyk2009book}
E.~D. Kolaczyk, \emph{Statistical Analysis of Network Data: Methods and
  Models}.\hskip 1em plus 0.5em minus 0.4em\relax New York, NY: Springer, 2009.

\bibitem{EmergingFieldGSP}
D.~Shuman, S.~Narang, P.~Frossard, A.~Ortega, and P.~Vandergheynst, ``The
  emerging field of signal processing on graphs: Extending high-dimensional
  data analysis to networks and other irregular domains,'' \emph{IEEE Signal
  Process. Mag.}, vol.~30, no.~3, pp. 83--98, Mar. 2013.

\bibitem{SandryMouraSPG_TSP13}
A.~Sandryhaila and J.~Moura, ``Discrete signal processing on graphs,''
  \emph{IEEE Trans. Signal Process.}, vol.~61, no.~7, pp. 1644--1656, Apr.
  2013.

\bibitem{SandryMouraSPG_TSP14Freq}
------, ``Discrete signal processing on graphs: Frequency analysis,''
  \emph{IEEE Trans. Signal Process.}, vol.~62, no.~12, pp. 3042--3054, June
  2014.

\bibitem{DongLaplacianLearning}
X.~Dong, D.~Thanou, P.~Frossard, and P.~Vandergheynst, ``Learning {L}aplacian
  matrix in smooth graph signal representations,'' \emph{arXiv preprint
  arXiv:1406.7842v2}, 2015.

\bibitem{MeiGraphStructure}
J.~Mei and J.~Moura, ``Signal processing on graphs: Estimating the structure of
  a graph,'' in \emph{IEEE Intl. Conf. Acoust., Speech and Signal Process.
  (ICASSP)}, 2015, pp. 5495--5499.

\bibitem{pasdeloup2016inferenceTSIPN16}
B.~Pasdeloup, V.~Gripon, G.~Mercier, D.~Pastor, and M.~G. Rabbat,
  ``Characterization and inference of weighted graph topologies from
  observations of diffused signals,'' \emph{arXiv preprint arXiv:1605.02569},
  2016.

\bibitem{Kalofolias2016inference_smoothAISTATS16}
V.~Kalofolias, ``How to learn a graph from smooth signals,'' in \emph{Intl.
  Conf. Artif. Intel. Stat. (AISTATS)}.\hskip 1em plus 0.5em minus 0.4em\relax
  J Mach. Learn. Res., 2016, pp. 920--929.

\bibitem{marques2016stationaryTSP16}
A.~G. Marques, S.~Segarra, G.~Leus, and A.~Ribeiro, ``Stationary graph
  processes and spectral estimation,'' \emph{arXiv preprint arXiv:1603.04667},
  2016.

\bibitem{perraudinstationary2016}
N.~Perraudin and P.~Vandergheynst, ``Stationary signal processing on graphs,''
  \emph{arXiv preprint arXiv:1601.02522}, 2016.

\bibitem{sporns2012book}
O.~Sporns, \emph{Discovering the Human Connectome}.\hskip 1em plus 0.5em minus
  0.4em\relax Boston, MA: MIT Press, 2012.

\bibitem{GLasso2008}
J.~Friedman, T.~Hastie, and R.~Tibshirani, ``Sparse inverse covariance
  estimation with the graphical lasso,'' \emph{Biostatistics}, vol.~9, no.~3,
  pp. 432--441, 2008.

\bibitem{Lake10discoveringstructure}
B.~M. Lake and J.~B. Tenenbaum, ``Discovering structure by learning sparse
  graph,'' in \emph{Annual Cognitive Sc. Conf.}, 2010, pp. 778 -- 783.

\bibitem{slawski2015estimation}
M.~Slawski and M.~Hein, ``Estimation of positive definite {M}-matrices and
  structure learning for attractive gaussian markov random fields,''
  \emph{Linear Algebra and its Applications}, vol. 473, pp. 145--179, 2015.

\bibitem{meinshausen06}
N.~Meinshausen and P.~Buhlmann, ``High-dimensional graphs and variable
  selection with the lasso,'' \emph{Ann. Stat.}, vol.~34, pp. 1436--1462, 2006.

\bibitem{pavez_laplacian_inference_icassp16}
E.~Pavez and A.~Ortega, ``Generalized {L}aplacian precision matrix estimation
  for graph signal processing,'' in \emph{IEEE Intl. Conf. Acoust., Speech and
  Signal Process. (ICASSP)}, Shanghai, China, Mar. 20-25, 2016.

\bibitem{BazerqueGeneNetworks}
X.~Cai, J.~A. Bazerque, and G.~B. Giannakis, ``Sparse structural equation
  modeling for inference of gene regulatory networks exploiting genetic
  perturbations,'' \emph{PLoS, Computational Biology}, Jun. 2013.

\bibitem{BainganaInfoNetworks}
B.~Baingana, G.~Mateos, and G.~B. Giannakis, ``Proximal-gradient algorithms for
  tracking cascades over social networks,'' \emph{IEEE J. Sel. Topics Signal
  Process.}, vol.~8, pp. 563--575, Aug. 2014.

\bibitem{Brovelli04Granger}
A.~Brovelli, M.~Ding, A.~Ledberg, Y.~Chen, R.~Nakamura, and S.~L. Bressler,
  ``Beta oscillations in a large-scale sensorimotor cortical network:
  directional influences revealed by granger causality,'' \emph{PNAS}, vol.
  101, p. 9849–9854, 2004.

\bibitem{Karanikolas_icassp16}
G.~V. Karanikolas, G.~B. Giannakis, K.~Slavakis, and R.~M. Leahy,
  ``Multi-kernel based nonlinear models for connectivity identification of
  brain networks,'' in \emph{IEEE Intl. Conf. Acoust., Speech and Signal
  Process. (ICASSP)}, Shanghai, China, Mar. 20-25, 2016.

\bibitem{shen2016kernelsTSP16}
Y.~Shen, B.~Baingana, and G.~B. Giannakis, ``Kernel-based structural equation
  models for topology identification of directed networks,'' \emph{arXiv
  preprint arXiv:1605.03122}, 2016.

\bibitem{Marks2011proteins}
D.~S. Marks, L.~J. Colwell, R.~Sheridan, T.~A. Hopf, A.~Pagnani, R.~Zecchina,
  and C.~Sander, ``Protein 3d structure computed from evolutionary sequence
  variation,'' \emph{PLoS ONE}, vol.~6, no.~12, p. e28766, 2011.

\bibitem{segarra2015graphfilteringTSP15}
S.~Segarra, A.~G. Marques, and A.~Ribeiro, ``Distributed linear network
  operators using graph filters,'' \emph{arXiv preprint arXiv:1510.03947},
  2015.

\bibitem{smola2003kernels}
A.~J. Smola and R.~Kondor, ``Kernels and regularization on graphs,'' in
  \emph{Learning theory and kernel machines}.\hskip 1em plus 0.5em minus
  0.4em\relax Springer, 2003, pp. 144--158.

\bibitem{chung1997spectral}
F.~Chung, \emph{Spectral Graph Theory}.\hskip 1em plus 0.5em minus 0.4em\relax
  American Mathematical Soc., 1997, vol.~92.

\bibitem{biyikougu2007laplacian}
T.~Biyiko{\u{g}}u, J.~Leydold, and P.~Stadler, \emph{Laplacian Eigenvectors of
  Graphs: Perron-Frobenius and Faber-Krahn Type Theorems}.\hskip 1em plus 0.5em
  minus 0.4em\relax Springer-Verlag Berlin Heidelberg, 2007.

\bibitem{segarra2015reconstruction}
S.~Segarra, A.~G. Marques, G.~Leus, and A.~Ribeiro, ``Reconstruction of graph
  signals through percolation from seeding nodes,'' \emph{IEEE Trans. Signal
  Process.}, vol.~64, no.~16, pp. 4363--4378, Aug 2016.

\bibitem{EUSIPCO_our_interp_2015}
------, ``Interpolation of graph signals using shift-invariant graph filters,''
  in \emph{European Signal Process. Conf. (EUSIPCO)}, 2015, pp. 210--214.

\bibitem{ssamar_distfilters_allerton15}
S.~Segarra, A.~G. Marques, and A.~Ribeiro, ``Distributed implementation of
  linear network operators using graph filters,'' in \emph{Allerton Conf. on
  Commun. Control and Computing}, 2015, pp. 1406--1413.

\bibitem{FeiziNetworkDeconvolution}
S.~Feizi, D.~Marbach, M.~Medard, and M.~Kellis, ``Network deconvolution as a
  general method to distinguish direct dependencies in networks,'' \emph{Nat
  Biotech}, vol.~31, no.~8, pp. 726--733, 2013.

\bibitem{candes_l0_surrogate}
E.~J. Candes, M.~B. Wakin, and S.~Boyd, ``Enhancing sparsity by reweighted
  $\ell_1$ minimzation,'' \emph{Journal of Fourier Analysis and Applications},
  vol.~14, pp. 877--905, Dec. 2008.

\bibitem{ChenDonohoBP}
S.~S. Chen, D.~L. Donoho, and M.~A. Saunders, ``Atomic decomposition by basis
  pursuit,'' \emph{SIAM Review}, vol.~43, no.~1, pp. 129--159, 2001.

\bibitem{zhang2013one}
H.~Zhang, M.~Yan, and W.~Yin, ``One condition for solution uniqueness and
  robustness of both l1-synthesis and l1-analysis minimizations,'' \emph{arXiv
  preprint arXiv:1304.5038}, 2013.

\bibitem{Ortega90}
J.~Ortega, \emph{Numerical Analysis: A Second Course}, ser. Classics in Applied
  Mathematics.\hskip 1em plus 0.5em minus 0.4em\relax Society for Industrial
  and Applied Mathematics, 1990.

\bibitem{bollobas1998random}
B.~Bollob{\'a}s, \emph{Random Graphs}.\hskip 1em plus 0.5em minus 0.4em\relax
  Cambridge University Press, 2001.

\bibitem{hagmann2008mapping}
P.~Hagmann, L.~Cammoun, X.~Gigandet, R.~Meuli, C.~J. Honey, V.~J. Wedeen, and
  O.~Sporns, ``Mapping the structural core of human cerebral cortex,''
  \emph{PLoS Biol}, vol.~6, no.~7, p. e159, 2008.

\bibitem{manning2008introduction}
C.~Manning, P.~Raghavan, and H.~Sch{\"u}tze, \emph{Introduction to Information
  Retrieval}.\hskip 1em plus 0.5em minus 0.4em\relax Cambridge University
  Press, 2008.

\end{thebibliography}

\end{document}